\documentclass[10pt,aps,prd,twocolumn,superscriptaddress,nofootinbib,nobibnotes,longbibliography,floatfix,groupedaddress]{revtex4-2}

\usepackage{bm}
\usepackage{mathtools,
amsmath,
amssymb,
amsfonts,
mathrsfs,
chngcntr}

\let\cc\corresponds
\let\corresponds\relax
\usepackage{mathabx}
\let\corresponds\cc

\usepackage[utf8]{inputenc}
\usepackage[T1]{fontenc}

\usepackage[dvipsnames]{xcolor}
\usepackage[unicode]{hyperref}
\hypersetup{colorlinks=true, citecolor=MidnightBlue,
            linkcolor=MidnightBlue, urlcolor=MidnightBlue, linktocpage=true}
\usepackage[normalem]{ulem}

\DeclareMathAlphabet{\mathpzc}{OT1}{pzc}{m}{it}
\definecolor{darkgreen}{rgb}{0.0, 0.6, 0.0}

\newcommand{\note}[1]{\text{\scshape\tiny{#1}}}

\newcommand{\ee}{\mathrm{e}}
\newcommand{\ii}{\mathrm{i}}
\newcommand{\dd}{\mathrm{d}}

\newcommand{\f}{\mathpzc{f}\!\!\mathpzc{f}}

\newcommand{\GN}{G_\note{N}}
\newcommand{\const}{\mathsf{const}}

\newcommand{\sst}{\sin^2\!\th}
\newcommand{\cct}{\cos^2\!\th}

\newcommand{\Q}{\mathcal{Q}}

\newcommand{\A}{\mathcal{A}}
\newcommand{\B}{\mathcal{B}}
\newcommand{\C}{\mathcal{C}}

\newcommand{\Hs}{\mathcal{H}}

\newcommand{\X}{\mathcal{X}}
\newcommand{\F}{\mathcal{F}}
\newcommand{\G}{\mathcal{G}}
\newcommand{\J}{\mathcal{J}}
\newcommand{\K}{\mathcal{K}}
\newcommand{\Sc}{\mathcal{S}}

\newcommand{\Ord}{\mathcal{O}}

\newcommand{\Eq}{\mathpzc{E}}
\newcommand{\Rel}{\mathpzc{R}}

\newcommand{\al}{\alpha}
\newcommand{\be}{\beta}

\newcommand{\de}{\delta}
\newcommand{\De}{\Delta}

\newcommand{\cep}{\varepsilon}
\newcommand{\ze}{\zeta}

\renewcommand\th{\theta}

\newcommand{\ka}{\kappa}
\newcommand{\la}{\lambda}
\newcommand{\La}{\Lambda}

\newcommand{\Sg}{\Sigma}
\newcommand{\cf}{\varphi}

\newcommand{\om}{\omega}
\newcommand{\Om}{\Omega}

\newcommand{\dl}{\partial}

\begin{document}

\author{Nicola Franchini}\email{franchini@apc.in2p3.fr}
\affiliation{Université Paris Cit\'e, CNRS, Astroparticule et Cosmologie,  F-75013 Paris, France}
\affiliation{CNRS-UCB International Research Laboratory, Centre Pierre Binétruy, IRL2007, CPB-IN2P3, Berkeley, CA 94720, USA}

\title{Slow rotation black hole perturbation theory}

\begin{abstract}
In this paper, we present a detailed analysis of first-order perturbations of the Kerr metric in the slow-rotation limit. We perform the calculation by perturbing the Schwarzschild metric plus up to second-order corrections in the spin in the Regge-Wheeler gauge. The apparent coupling between different angular momentum axial-led and polar-led modes can be removed by suitably combining the perturbation equations and projecting them onto spin-weighted spherical harmonics. In this way, we derive the corrections to the Regge-Wheeler and the Zerilli equations up to second-order in the spin. We show that the two potentials remain isospectral as in the non-rotating limit. However, it is easy to demonstrate it only for a precise choice of the tortoise coordinate. The isospectrality with slow-rotating Teukolsky equation is also verified. We discuss the main implication of this result for the problem of vacuum metric reconstruction, providing the transformation rule between slow-spinning Teukolsky variables and metric perturbations. The existence of this relation leaves us with the conjecture that a resummation of the expansion in the spin is possible, leading to two decoupled differential equations for perturbations of the Kerr metric.
\end{abstract}

\maketitle

\section{Introduction}\label{sec:introduction}
Black hole perturbation theory (BHPT) is the branch of gravitational physics that studies the response of black holes (BHs) to small generic fluctuations of the spacetime. It was initially developed in the 1957 breakthrough work by Regge and Wheeler~\cite{Regge:1957td}, who for the first time obtained the first-order equations for a perturbation with axial parity on top of a Schwarzschild BH. Zerilli then in 1970 obtained a similar equation for even parity perturbations~\cite{Zerilli:1970se}. Few years later, Teukolsky managed to derive an equation that governs linear perturbations on top of a Kerr BH, by implementing a different formalism based on the perturbations of the curvature~\cite{Teukolsky:1973ha}. These three milestones are still nowadays the basic equations for the analysis of small perturbations of rotating and non-rotating BH in general relativity (GR). Notable examples of its applications are the study of the quasi-normal modes of BHs~\cite{Berti:2009kk,Franchini:2023eda} and the waveform generation of black hole binaries with very large mass ratio~\cite{Drasco:2005kz,Hughes:2021exa,Islam:2022laz,Wardell:2021fyy}.

The Regge-Wheeler and the Zerilli equations have both the form of a Schr\"odinger equation, but their effective potentials have different analytical expression. Nevertheless, Chandrasekhar found that one can transform the Regge-Wheeler equation into the Zerilli equation and back~\cite{Chandrasekhar:1975nkd,Chandrasekhar:1985kt}. The existence of this transformation confirms the {\it isospectrality} of the two potentials, which means that their spectrum of quasi-normal modes is completely equivalent~\cite{Chandrasekhar:1975zza}. In addition, Chandrasekhar found that the Regge-Wheeler and the Zerilli equations can be related, with a slightly more complex transformation, to the non-rotating limit of the Teukolsky equation~\cite{Chandrasekhar:1985kt}, also known as the Bardeen-Press equation~\cite{Bardeen:1973xb}. This latter result by Chandrasekhar confirms a somewhat expected property: the spectrum of oscillations of a BH does not depend on the perturbation scheme used to calculate it. A direct perturbation of the spacetime metric must lead to an equivalent result as if one performs a different perturbation scheme, like for the derivation of the Teukolsky equation.

The transformations derived by Chandrasekhar not only prove the isospectrality between the three different equations, but also provides the transformation rules to move from one to another. While this result in the non-rotating case is rather immediate, having an exact formula that links the solution of the Teukolsky equation to the small perturbations of the Kerr metric is a much more involved problem commonly known as ``metric reconstruction''. Such reconstruction appears to be necessary in all those problems of BHPT which involve ``perturbations of perturbations'': second-order perturbations of Kerr BHs in self-force computations~\cite{Pound_2021} or in the case of modified Teukolsky equations in alternative theories of gravity~\cite{Li:2022pcy,Hussain:2022ins,Cano:2023tmv}

The main issue in the rotating cases is that an equivalent of the Regge-Wheeler/Zerilli equation cannot be found and the perturbations of a rotating BH can only be understood via the Teukolsky equation. The procedure to obtain it uses a decomposition of the spacetime for which it is not straightforward to trace it back to the actual perturbations of the metric. Nevertheless, there are techniques that aim precisely to obtain the metric perturbations starting from the Teukolsky variables and they differ among themselves upon the gauge choice~\cite{Pound_2021}. In the radiation gauge, one possibility is to integrate over the Hertz potentials~\cite{Chrzanowski:1975wv} (see also~\cite{Whiting:2005hr} for an overview of the technique and possible implementations). It is worth noting that recent progress showed an alternative formulation based on the calculations of Chandrasekhar~\cite{Chandrasekhar:1985kt} which avoids the use of Hertz potentials~\cite{Loutrel:2020wbw}. Another alternative is to perform the reconstruction in the Lorentz gauge~\cite{Dolan:2021ijg}.

In this paper, we analyse the problem of metric reconstruction in the Regge-Wheeler gauge extending the results of~\cite{Lousto:2005xu} for a vacuum spacetime up to second-order in the spin. Indeed, it is known since the early works of Kojima on the perturbation of slowly-rotating neutron stars~\cite{Kojima:1992ie,1993ApJ...414..247K} that first-order corrections in the spin maintain the same structure of the perturbation equations as in the non-rotating case. The effect of the spin on BH linear perturbations is just to modify the Regge-Wheeler and Zerilli potential~\cite{Lousto:2010qx,Nakano:2010kv,Pani:2013pma}. The slow-spin expansion is particularly relevant in those cases where the perturbative approach to the full rotating problem is not possible, like in alternative theories of gravity where fully rotating solutions are not known analytically~\cite{Cano:2021myl,Pierini:2021jxd,Wagle:2021tam,Srivastava:2021imr,Pierini:2022eim} or when exotic matter fields coupled to gravity do not lead to an evident separation of the variables in the perturbation equations~\cite{Pani:2012bp,Pani:2012vp}.

The outline of the paper is the following. In section~\ref{sec:metric&perturbation} we present the slow-spinning Kerr metric and the perturbation scheme, followed by a revision of the method of Kojima to perform the separation of the equations. The method is extended up to second-order in the spin, as it was done in~\cite{Pierini:2022eim}, and we show how to manipulate the equations in order to find a correction in the spin for the Regge-Wheeler and the Zerilli potentials. In section~\ref{sec:isospectrality} we discuss how the property of isospectrality remains satisfied at second-order in the spin, as well as with respect to the slow-spinning Teukolsky equation as shown in section~\ref{sec:iso_teuk}. We discuss the main results of the paper in section~\ref{sec:discussion}, where we summarize the steps necessary to perform the metric reconstruction, we provide a formula that generates the corrections due to the spin to the effective potential of the Regge-Wheeler and the Zerilli equations, and we conjecture the existence of two fully rotating versions of those equations. Finally, in section~\ref{sec:conclusions}, we argue how the results of this paper could address some currently open problems and which future directions can be investigated.

Throughout the paper we use mostly minus signature $(+,-,-,-)$. This choice was made to be in conformity of the notation of~\cite{Chandrasekhar:1985kt}. Moreover, we use units such that $c=\GN=1$.

\section{Slowly-rotating Regge-Wheeler and Zerilli equations}\label{sec:metric&perturbation}
\subsection{The perturbation scheme}
With $g_{ab}^\note{K}$ we denote the Kerr metric in Boyer-Lindquist coordinates $x^a = (t,r,\th,\cf)$, whose line element $\dd s^2_\note{K} = g_{ab}^\note{K} \dd x^a \dd x^b $ is
\begin{equation}\label{eq:Kerr}
\begin{split}
    \dd s^2_\note{K} = & \left( 1 - \frac{r}{\Sg} \right) \dd t^2 + \frac{2 a r \sst}{\Sg}\dd t \dd \cf - \frac{\Sg}{\De}\dd r^2 \\ & - \Sg\,\dd\th^2 -  \sst \left( r^2 + a^2 + \frac{a^2 r \sst}{\Sg}\right) \dd\cf^2 ,
\end{split}
\end{equation}
where $\De = r^2 - r + a^2$ and $\Sg = r^2 + a^2\cct$. Without loss of generality we chose units such as $M=1/2$, being $M$ the mass parameter of the Kerr metric. 

Since we are interested in the slow-rotation BHPT, we need to perform a double perturbation scheme: the usual linear perturbation for which we introduce a formal bookkeeping parameter $\cep$, as well as an expansion in the spin parameter $a$, which for the purposes of this paper we truncate at second-order. In this way, we are formally considering the following ansatz for the metric
\begin{equation}\label{eq:linearized_ansatz}
    g_{ab} = g^\note{srK}_{ab} + \cep h_{ab},
\end{equation}
with $g^\note{srK}_{ab}$ being the expansion of the metric in Eq.~\eqref{eq:Kerr} up to second-order in $a$
\begin{align}
    g^\note{srK}_{tt} = & f_0 + a^2\frac{\cct}{r^3} , \qquad
    g^\note{srK}_{t\cf} = \frac{2 a \sst}{r} , \\
    g^\note{srK}_{rr} = &  - \frac{1}{f_0} + a^2 \frac{1-f_0\cct}{r^2 f_0^2} , \\
    g^\note{srK}_{\th\th} = &  - r^2 - a^2\cct , \\
    g^\note{srK}_{\cf\cf} = &  - \sst \left[ r^2 + a^2 \frac{r+\sst}{r}\right] ,
\end{align}
being $f_0 = 1 - 1/r$.
If we treat $g^\note{srK}_{ab}$ as a correction to the Schwarzschild metric, we can express the perturbation metric $h_{ab}$ in the Regge-Wheeler gauge, which decomposes in a sum of axial $h_{ab}^{(-)}$ and polar $h_{ab}^{(+)}$ contributions. Their explicit expression respectively reads as
\begin{align}
    h_{ab}^{(-)} & \dd x^a \dd x^b  = 2\left[ h_0^{\ell m}(r) \dd t + h_1^{\ell m}(r) \dd r \right] \notag \\
    & \times \left[ S_\th^{\ell m}(\th,\cf) \dd \th + S_\cf^{\ell m}(\th,\cf) \dd \cf \right]\ee^{\rho t}, \\
    h_{ab}^{(+)} & \dd x^a \dd x^b  = \Bigg[f_0 H_0^{\ell m}(r) \dd t^2 + 2 H_1^{\ell m }(r) \dd t \dd r \notag \\
    & +\frac{H_2^{\ell m}(r)}{f_0} \dd r^2 + K^{\ell m}(r) \dd\Om^2 \Bigg] Y^{\ell m}(\th,\cf) \ee^{\rho t},
\end{align}
where we performed a Fourier decomposition in modes of frequency $\om = -\ii \rho$, the functions $Y^{\ell m}(\th, \cf)$ are scalar spherical harmonics and
\begin{equation}
    \left( S_\th^{\ell m}, S_\cf^{\ell m} \right) = \left( - \frac{Y_{,\cf}^{\ell m}}{\sin\th} ,  \sin\th \, Y_{,\th}^{\ell m}\right) ,
\end{equation}
where the comma followed by a coordinate stands for partial derivative with respect to that variable.
It must be noticed how the functions $h_0^{\ell m}$, $h_1^{\ell m}$, $H_0^{\ell m}$, $H_1^{\ell m}$, $H_2^{\ell m}$, $K^{\ell m}$ are free radial functions, and we will see how to relate them to their non-rotating counterpart.
We can plug the linearized ansatz~\eqref{eq:linearized_ansatz} into the Einstein equations and solve them with a double-perturbation scheme in $\cep$ and $a$. In total there are ten equations which we schematically dub
\begin{equation}\label{eq:LinEq}
    \de\Eq_{ab} \equiv R_{ab}^{(1)} = 0,
\end{equation}
where we defined the linear perturbation of the Ricci tensor as $R_{ab} = R_{ab}^{(0)} + \cep R_{ab}^{(1)}$. In principle, the presence of the spin introduces an angular dependency in $\th$, which makes the equations non-manifestly separable. In the next section, we show a scheme, which is valid up to second-order in $a$, but it is in principle extendable to arbitrary order, that allows to separate the equations, and obtain two master radial equations. 

\subsection{Decoupling the equations}\label{sec:angular_separation}

Here, we apply the scheme for the decoupling of the equations in the slow-spin expansion as in~\cite{Kojima:1992ie}, and we extend it up to the second-order in the spin, as it was done in~\cite{Pierini:2022eim}. The ten equations \eqref{eq:LinEq} divide in three different groups, according to their functional form. From now on, to avoid cluttering of indices we omit the superscript index $m$, since different values are always decoupled in the equations. The first group schematically reads as
\begin{equation}\label{eq:firstgroup}
\begin{split}
\delta\Eq_{(i)}\equiv & \left(A^{(i)}_{0,\ell}+A^{(i)}_{1,\ell}\cos\th+A^{(i)}_{2,\ell}\cos\th^2\right)Y^\ell \\
& +\left(B^{(i)}_{1,\ell} +B^{(i)}_{2,\ell}\cos\th\right)\sin\th\,Y^\ell_{,\theta}=0,
\end{split}
\end{equation}
where the index $i$ runs from 0 to 3 and corresponds to $\delta\Eq_{tt}=0$, $\delta\Eq_{tr}=0$, $\delta\Eq_{rr}=0$ and $\delta\Eq_{\theta\theta}+\delta\Eq_{\varphi\varphi}/\sst=0$, respectively. The second group reads
\begin{subequations}\label{eq:secondgroup}
\begin{equation}\label{eq:secondgroup1}
\begin{split}
& \delta\Eq_{(j\theta)} \equiv 
\left(\sum_{n=0}^2 \al_{n,\ell}^{(j)}\cos^n\!\th +\widetilde{\alpha}^{(j)}_{2,\ell}\sst\right)Y^\ell_{,\theta} \\
& - \left(\sum_{n=0}^2 \beta^{(j)}_{n,\ell} \cos^n\!\th + \widetilde{\beta}^{(j)}_{2,\ell}\sst\right)\frac{Y^\ell_{,\varphi}}{\sin\th} \\
& + \left(\eta^{(j)}_{1,\ell}+\eta^{(j)}_{2,\ell}\cos\th\right)\sin\th\,Y^\ell + \chi^{(j)}_{1,\ell} \sin\th\,W^\ell \\
& + \left(\xi^{(j)}_{1,\ell}+\xi^{(j)}_{2,\ell}\cos\th\right)X^\ell =0
,
\end{split}
\end{equation}
\begin{equation}\label{eq:secondgroup2}
\begin{aligned}
& \delta\Eq_{(j\varphi)}\equiv \left(\sum_{n=0}^2 \beta^{(j)}_{n,\ell} \cos^n\!\th -\widetilde{\beta}^{(j)}_{2,\ell}\sst\right)Y^\ell_{,\theta} \\
& + \left(\sum_{n=0}^2 \al_{n,\ell}^{(j)}\cos^n\!\th - \widetilde{\alpha}^{(j)}_{2,\ell}\sst\right)\frac{Y^\ell_{,\varphi}}{\sin\th} \\
& + \left(\zeta^{(j)}_{1,\ell}+\zeta^{(j)}_{2,\ell}\cos\th\right)\sin\th\,Y^\ell +\chi^{(j)}_{1,\ell}\,X^\ell \\
& - \left(\xi^{(j)}_{1,\ell}+\xi^{(j)}_{2,\ell}\cos\th\right)\sin\th\,W^\ell=0,
\end{aligned}
\end{equation}
\end{subequations}
where $j=0,1$ corresponds to $\delta\Eq_{t\theta}=0$ and $\delta\Eq_{r\theta}=0$ for the first equation and $\delta\Eq_{t\varphi}=0$ and $\delta\Eq_{r\varphi}=0$ for the second equation. The symbols $X^\ell$ and $W^\ell$ are related to the spin-2 spherical harmonics and are defined as
\begin{align}\label{eq:X}
X^\ell & = 2Y^\ell_{,\theta\varphi}-2\cot\th\,Y^\ell_{,\varphi} , \\
\label{eq:W}
W^\ell & = Y^\ell_{,\theta\theta}-\cot\th\,Y^\ell_{,\theta}-\frac{Y^\ell_{,\varphi\varphi}}{\sst}.
\end{align}
Finally, the third group reads
\begin{subequations}\label{eq:thirdgroup}
\begin{equation}\label{eq:thirdgroup1}
\begin{aligned}
\delta\Eq_{(\theta\varphi)}\equiv & \left(f_{1,\ell}+f_{2,\ell}\cos\th\right)\sin\th\,Y^\ell_{,\theta} \\
+ & \left(g_{1,\ell}+g_{2,\ell}\cos\th\right)Y^\ell_{,\varphi} \\ 
+ & h_{2,\ell} \sst\,Y^\ell + \left(j_{0,\ell}+j_{2,\ell}\cct\right)\frac{X^\ell}{\sin\th} \\
+ & \left(k_{0,\ell}+k_{2,\ell}\cct\right)W^\ell=0,
\end{aligned}
\end{equation}
\begin{equation}\label{eq:thirdgroup2}
\begin{aligned}
\delta\Eq_{(-)}\equiv & \left(g_{1,\ell}+g_{2,\ell}\cos\th\right)\sin\th\,Y^\ell_{,\theta} \\
- & \left(f_{1,\ell}+f_{2,\ell}\cos\th\right)Y^\ell_{,\varphi} \\ 
+ &\widetilde{h}_{2,\ell} \sst\,Y^\ell - \left(k_{0,\ell}+k_{2,\ell}\cct\right)\frac{X^\ell}{\sin\th} \\
+ & \left(j_{0,\ell}+j_{2,\ell}\cct\right)W^\ell=0,
\end{aligned}
\end{equation}
\end{subequations}
with $\delta\Eq_{(-)}=\delta\Eq_{\theta\theta}-\delta\Eq_{\varphi\varphi}/\sst=0$. With this schematic representation of the equations, the functions $A_{n,\ell}^{(i)}$, $B_{n,\ell}^{(i)}$, $\al_{n,\ell}^{(j)}$, $\be_{n,\ell}^{(j)}$, $\widetilde{\al}_{n,\ell}^{(j)}$, $\widetilde{\be}_{n,\ell}^{(j)}$, $\eta_{n,\ell}^{(j)}$, $\xi_{n,\ell}^{(j)}$, $\chi_{n,\ell}^{(j)}$, $f_{n,\ell}$, $g_{n,\ell}$, $h_{n,\ell}$, $\widetilde{h}_{n,\ell}$, $k_{n,\ell}$, $j_{n,\ell}$ are purely radial functions, which contain combinations of the metric perturbation functions and their radial derivative. We labelled each function with the index $n$ such that it contains at least $\Ord(a)^n$ terms. Their explicit expression can be found in the supplemental material. To separate radial and angular components from the equations, we make use of the completeness relation of spherical harmonics
\begin{align}\label{eq:completeness}
    & \int \dd\Om \, Y^\ell Y^{*\ell'} = \de^{\ell \ell'},
\end{align}
where $*$ denotes complex conjugation, by the fact that spherical harmonics satisfy the equation
\begin{equation}\label{eq:spherical_hoarmonics}
    Y_{,\th\th}^\ell + \cot\th \, Y_{,\th}^{\ell} + \frac{ Y_{,\cf\cf}^\ell}{\sst} = - \ell(\ell+1) Y^\ell ,
\end{equation}
as well as the following relations among combinations of spherical harmonics and trigonometric functions
\begin{align}
\label{eq:cosY}
    \cos\th \, Y^\ell & = \Q_{\ell+1} Y^{\ell+1} + \Q_\ell Y^{\ell-1} , \\
\label{eq:sinY}
    \sin\th \, Y_{,\th}^\ell & = \ell \, \Q_{\ell+1} Y^{\ell+1} -(\ell+1) \Q_\ell Y^{\ell-1} ,
\end{align}
where we defined $\Q_{\ell} = \sqrt{(\ell^2 - m^2)/(4\ell^2-1)}$.
One can repeatedly apply the formulas~\eqref{eq:cosY}--\eqref{eq:sinY} to find similar expression for $\cos^n\!\th\, Y^\ell$ and $\cos^n\!\th\, \sin\th \, Y_{,\th}^\ell$. The useful ones that we used for our calculation are shown in Appendix~\ref{app:integrals}.
The equations of the three groups can be separated by taking suitable linear combinations of the equations and integrating them over the 2-sphere. The decoupled equations are obtained in the following schematic form:
\begin{widetext}
\begin{align}
    \mathcal{E}^\note{I}_{(i+)} \equiv \int \dd \Om \, \de\Eq_{(i)} Y^{*\ell'} & = \sum_{n=0}^2 \left[ \C_n A_{n,\ell}^{(i)}  + \Sc_n B_{n,\ell}^{(i)} \right]  ,\\
    \mathcal{E}^\note{II}_{(j+)} \equiv \int \dd \Om \left( \de \Eq_{(j\th)} Y_{,\th}^{*\ell'} + \frac{\de \Eq_{(j\cf)} Y_{,\cf}^{*\ell'}}{\sin\th}\right) & = \sum_{n=0}^2 \left[ \A_n \al_{n,\ell}^{(j)} + \B_n \be_{n,\ell}^{(j)} \right] + \widetilde{\A}_2 \widetilde{\al}_{2,\ell}^{(j)} + \widetilde{\B}_2 \widetilde{\be}_{2,\ell}^{(j)} \notag \\
    & + \sum_{n=1}^{2} \left[ \widebar{\Sc}_{n} \eta_{n,\ell}^{(j)} + \ii m \, \C_{n-1} \ze_{n,\ell}^{(j)} + \X_{n-1} \xi_{n,\ell}^{(j)} + \widebar{\X}_{n-1} \chi_{n,\ell}^{(j)}  \right] , \\
    \mathcal{E}^\note{II}_{(j-)} \equiv  \int \dd \Om \left( \de \Eq_{(j\cf)} Y_{,\th}^{*\ell'} - \frac{\de \Eq_{(j\th)} Y_{,\cf}^{*\ell'}}{\sin\th}\right) & = \sum_{n=0}^2 \left[ - \A_n \be_{n,\ell}^{(j)} + \B_n \al_{n,\ell}^{(j)} \right]  + \widetilde{\A}_2 \widetilde{\be}_{2,\ell}^{(j)} - \widetilde{\B}_2 \widetilde{\al}_{2,\ell}^{(j)} \notag \\
    & + \sum_{n=1}^{2} \left[ \widebar{\Sc}_{n} \ze_{n,\ell}^{(j)} - \ii m \, \C_{n-1} \eta_{n,\ell}^{(j)} + \X_{n-1} \chi_{n,\ell}^{(j)} - \widebar{\X}_{n-1} \xi_{n,\ell}^{(j)}  \right]  , \\
    \mathcal{E}^\note{III}_{(+)} \equiv \int \dd \Om \left( \de \Eq_{(-)} W^{*\ell'} + \frac{\de \Eq_{(\th\cf)} X^{*\ell'}}{\sin\th}\right) & = \sum_{n=0}^1 \left[ \F_n f_{n+1,\ell} + \G_n g_{n+1,\ell} + \J_{2n} j_{2n,\ell} + \K_{2n} k_{2n,\ell} \right] + \widebar{\Hs} h_{2,\ell} + \Hs \widebar{h}_{2,\ell} , \\
    \mathcal{E}^\note{III}_{(-)} \equiv \int \dd \Om \left( \de \Eq_{(\th\cf)} W^{*\ell'} - \frac{\de \Eq_{(-)} X^{*\ell'}}{\sin\th}\right) & = \sum_{n=0}^1 \left[ \G_n f_{n+1,\ell} - \F_n g_{n+1,\ell} + \J_{2n} k_{2n,\ell} - \K_{2n} j_{2n,\ell} \right] - \widebar{\Hs}\, \widebar{h}_{2,\ell} + \Hs h_{2,\ell} .
\end{align}
\end{widetext}
We labelled the equations with a Roman number that reminds from which group they have been obtained, and with a $(+)/(-)$ the equations whose limit $a \to 0$ contains only polar/axial quantities. We refer to the first/second group of equations as to polar/axial-led, respectively. The operators $\C_n$, $\Sc_n$, $\A_n$, $\B_n$, $\widetilde{\A}_2$, $\widetilde{\B}_2$, $\widebar{\Sc}_n$, $\X_n$, $\widebar{\X}_n$, $\F_n$, $\G_n$, $\J_{2n}$, $\K_{2n}$, $\Hs$ and $\widebar{\Hs}$ are integrals which mix modes with different angular momentum $\ell$, and they are explicitly provided in Appendix~\eqref{app:integrals}.

The general structure of all the radial equations obtained with this procedure is
\begin{align}\label{eq:splitting}
    \mathcal{E}_\ell = \mathcal{P}^{\ell} + a \widebar{\mathcal{P}}^{\ell\pm1} + a^2 \mathcal{P}^{\ell\pm2},
\end{align}
where $\mathcal{P}$ refers to a combination of the functions, and their derivatives for a given parity, whereas $\widebar{\mathcal{P}}$ are combinations of functions, and their derivatives, of opposite parity. The $\ell$ label signals that functions of a chosen parity of angular momentum $\ell$ couple at least at first-order in the spin with functions of opposite parity and angular momentum $\ell\pm1$ and  at least at second-order in the spin with functions of same parity and angular momentum $\ell\pm2$. The spin factor outside each component signals just the minimum order at which the modification is entering. 

Let us notice that thanks to the linearity of the equations one can combine different equations and their radial derivatives in such a way that the structure denoted in equation~\eqref{eq:splitting} does not change. If we have any two equations $\mathcal{E}_{1}^\ell = 0$ and $\mathcal{E}_{2}^\ell = 0$ with the same parity in the non-spinning limit, as well as an equation $\widebar{\mathcal{E}}_{3}^\ell = 0$ with different parity in the non-spinning limit the structure~\eqref{eq:splitting} up to second-order with the spin is preserved if:
\begin{itemize}
    \item one takes a linear combination of $\mathcal{E}_{1}^\ell$, $\mathcal{E}_{2}^\ell$ as well as their radial derivatives;
    \item one takes a linear combination of $\mathcal{E}_{1}^\ell$, $a \widebar{\mathcal{E}}_{3}^{\ell\pm1}$ as well as their radial derivatives;
    \item one takes a linear combination of $\mathcal{E}_{1}^\ell$, $a^2\mathcal{E}_{2}^{\ell\pm2}$ as well as their radial derivatives.
\end{itemize}
We will use these three transformations extensively to drastically simplify the ten equations.

Finally, let us notice that the terms $\mathcal{P}_{\ell}$, $\widebar{\mathcal{P}}_{\ell\pm1}$ and $\mathcal{P}_{\ell\pm2}$ appearing in equation~\eqref{eq:splitting} have the following structure
\begin{align}
    & \mathcal{P}^{\ell} = \bigg[ A_0^\ell + a m A_1^\ell \notag \\
    & \quad + a^2 \left( A_2^\ell + m^2 \widebar{A}_2^\ell + \widetilde{A}_2^\ell \Q_{\ell+1}^2 + \widetilde{A}_2^{-\ell-1} \Q_{\ell}^2 \right) \bigg] f_\ell , \\
    & \widebar{\mathcal{P}}^{\ell\pm1} = \Q_{\ell+1} \left( B_1^{\ell} + a m B_2^{\ell} \right) \widebar{f}_{\ell+1} \notag \\
    & \qquad \; + \Q_\ell \left( B_1^{-\ell-1} + a m B_2^{-\ell-1} \right) \widebar{f}_{\ell -1}, \\
    & \mathcal{P}^{\ell\pm2} = \Q_{\ell+1} \Q_{\ell+2} C_2^{\ell} f_{\ell+2} + \Q_{\ell-1} \Q_\ell C_2^{-\ell-1} f_{\ell -2}, 
\end{align}
where with the symbol $f_\ell$ we refer to any of the perturbation function $h_0^\ell$, $h_1^\ell$, $H_0^\ell$, $H_1^\ell$, $H_2^\ell$, $K^\ell$ (and their derivatives) of a given parity, whereas $\widebar{f}$ stands for the same functions, but opposite parity, and $A_0^\ell$, $A_1^\ell$, $A_2^\ell$, $\widebar{A}_2^\ell$, $\widetilde{A}_2^\ell$, $B_1^\ell$, $B_2^\ell$ and $C_2^\ell$  are functions of $r$ and $\ell$ only. 
In this way, we completely determined how the index $m$ enters in the equations, and it is clear that different values of $m$ never couple to each other. In the next section we show how to re-define the perturbation variables such that they satisfy differential equations where also the coupling between different $\ell$ is removed.

\subsection{Spin corrections to the Regge-Wheeler and Zerilli equations}\label{subsec:RWZequations}
Let us start by denoting the polar-led equations as $Z_1 = \mathcal{E}^\note{I}_{(0+)}$, $Z_2 = \mathcal{E}^\note{I}_{(1+)}$, $Z_3 = \mathcal{E}^\note{I}_{(2+)}$, $Z_4 = \mathcal{E}^\note{I}_{(3+)}$, $Z_5 = \mathcal{E}^\note{II}_{(0+)}$, $Z_6 = \mathcal{E}^\note{II}_{(1+)}$, $Z_7 = \mathcal{E}^\note{III}_{(+)}$ and the axial-led equations as $Q_1 = \mathcal{E}^\note{II}_{(0-)}$, $Q_2 = \mathcal{E}^\note{II}_{(1-)}$, $Q_3 = \mathcal{E}^\note{III}_{(-)}$. We now show that these 10 equations are not independent, and how they can be recast into two independent equations which generalize the Regge-Wheeler and the Zerilli equation up to second-order in the spin.

Let us revise the derivation of the two equations in the limit $a=0$. From $Q_2 = 0$, we can find an expression for $\dl_r h_0^{\ell}$, while from $Q_3=0$ we obtain an expression for $\dl_r h_1^{\ell}$. It is straightforward to check that $Q_1=0$ is automatically satisfied, by taking combinations of $Q_2$, $Q_3$ and their derivatives. At this point, one can define
\begin{align}
\label{eq:RW_redef_h01}
    h_1^{\ell} (r) & = \frac{r}{f_0} \Phi_{(-)}^{\ell} (r), \qquad
    h_0^{\ell} (r) = \frac{f_0}{\rho} \dl_r\left[ r \Phi_{(-)}^{\ell} (r) \right].
\end{align}
By inserting these expressions into, {\it e.g.}, the equation $Q_2 = 0$, one finds that the function $\Phi_{(-)}^{\ell}$ satisfies the so-called Regge-Wheeler equation~\cite{Regge:1957td}
\begin{equation}
    \frac{\dd^2 \Phi_{(-)}^{\ell}}{\dd r_{*,0}^2} - \left( \rho^2 + V_{(-),0}^{\ell} \right) \Phi_{(-)}^{\ell} = 0 ,
\end{equation}
where $\dd r_{*,0} = \dd r / f_0$ is the Schwarzschild tortoise coordinate, and the Regge-Wheeler potential reads
\begin{equation}
    V_{(-),0}^{\ell} = f_0 \left[ \frac{\ell(\ell+1)}{r^2} - \frac{3}{r^3} \right].
\end{equation}

On the polar side, we can solve $Z_7=0$ to find an algebraic expression for $H_2^{\ell}$, then the combined solution to $Z_5=Z_2=Z_6=0$ leads to an expression for $\dl_r H_1^{\ell}$, $\dl_r K^{\ell}$, $\dl_r H_0^{\ell}$, respectively. The combination of the former results inserted into $Z_4=0$ leads to an algebraic expression for $H_0^{\ell}$. Again, with some algebraic manipulation one can show that $Z_1 = Z_3 = 0$ are automatically satisfied. The definition of the Zerilli function is made through 
\begin{align}
\label{eq:Zer_redef_K}
    K^{\ell} & = \left[ \frac{\la -2}{2r} - \frac{3 f_0}{r(3+\la r)} \right] \Phi_{(+)}^{\ell} + f_0 \dl_r \Phi_{(+)}^{\ell}, \\
\label{eq:Zer_redef_H1}
    H_1^{\ell} & = \left[ \frac{2r-3}{2rf_0} - \frac{3}{3+\la r}\right] \rho \Phi_{(+)}^{\ell} + \rho r \dl_r \Phi_{(+)}^{\ell}, 
\end{align}
where $\la = \ell(\ell+1) -2$
By inserting these expressions either into, {\it e.g.}, $Z_5 = 0$, one obtains the so-called Zerilli equation~\cite{Zerilli:1970se}
\begin{equation}
    \frac{\dd^2 \Phi_{(+)}^{\ell}}{\dd r_{*,0}^2} - \left( \rho^2 + V_{(+),0}^{\ell} \right) \Phi_{(+)}^{\ell} = 0 ,
\end{equation}
with the Zerilli potential being
\begin{equation}
    V_{(+),0}^{\ell} = f_0 \left[ \frac{\ell(\ell+1)}{r^2} - \frac{3}{r^3}\frac{r^2\la(\la+4)+6r-3}{(3+r \la)^2} \right].
\end{equation}

Let us now turn back to the full problem. One can formally perform the same passages to obtain second-order in $a$ expressions for $\dl_r h_0^{\ell}$ and $\dl_r h_1^{\ell}$ in the axial-led sector and for $\dl_r H_1^{\ell}$, $\dl_r K^{\ell}$, $\dl_r H_0^{\ell}$, $H_2^{\ell}$ and $H_0^{\ell}$ in the polar-led sector. The structure of these expressions is the same of Eq.~\eqref{eq:splitting}, because it is obtained by repeated use of the three combination rules enumerated in the previous section. Thus, for each given equation of a given parity, one can use the expressions found for the functions of opposite parity and $\ell\pm 1$ up to first-order in the spin and for the functions of same parity and $\ell\pm2$ up to zeroth order in the spin. The overall outcome is very similar to the spinless case, as also here $Q_1 = Z_1 = Z_3 = 0$ are automatically satisfied up to $\Ord(a^2)$. 

In order to obtain a generalization of the RW and the Zerilli equations we propose a modified redefinition of equations~\eqref{eq:RW_redef_h01} and~\eqref{eq:Zer_redef_K}--\eqref{eq:Zer_redef_H1} that takes into account couplings with functions of different angular momentum are introduced at each order in the spin. Given that all the equations can be written just in terms of $h_0^{\ell}$, $h_1^{\ell}$, $K^{\ell}$ and $H_1^{\ell}$, we guess the following ansatz for the redefinition of variables, based on the transformation rules that maintain the structure of the equation~\eqref{eq:splitting} unchanged\footnote{In principle, one should be able to express the functions $h_1$, $h_0$, $K$ and $H_1$ only in terms of the functions $\Phi_{(\pm)}$ with angular momentum $\ell\pm0,1,2$. However, since the transformations are invertible at each order in $a$, and they all have the form of equation~\eqref{eq:splitting}, we find that the explicit expression of the coefficients is more compact if expressed in terms of the original metric functions (see appendix~\ref{app:coefficients} for linear coefficients and supplemental material for quadratic ones).}
\begin{equation}
\begin{split}
\label{eq:RW_redef_h1_2nd_order}
    h_1^{\ell} & = \frac{r}{f_0} \Phi_{(-)}^{\ell} + \frac{\ii a m}{\rho} c_1^{\ell} \Phi_{(-)}^{\ell} + a^2 d_1^{\ell} \Phi_{(-)}^{\ell} \\
                  & + a \bigg[ \Q_\ell \left( s_1^{-\ell-1} H_1^{\ell-1} + t_1^{-\ell-1} K^{\ell -1} \right) \\
                  & \qquad + \Q_{\ell+1} \left( s_1^{\ell} H_1^{\ell+1} + t_1^{\ell} K^{\ell +1} \right) \bigg] \\
                  & + a^2 \bigg[ \Q_{\ell-1} \Q_\ell \left( u_1^{-\ell-1} h_1^{\ell-2} + v_1^{-\ell-1} h_0^{\ell-2} \right)  \\
                  & \qquad + \Q_{\ell+1} \Q_{\ell+2} \left( u_1^{\ell} h_1^{\ell+2} + v_1^{\ell} h_0^{\ell+2} \right) \bigg],
\end{split}
\end{equation}
\begin{equation}
\begin{split}
\label{eq:RW_redef_h0_2nd_order}
    h_0^{\ell} & = \frac{f_0}{\rho} \dl_r\left( r \Phi_{(-)}^{\ell} \right) + \frac{\ii a m}{\rho} c_0^{\ell} \Phi_{(-)}^{\ell}+ a^2 d_0^{\ell} \Phi_{(-)}^{\ell} \\
                  & + a \bigg[ \Q_\ell \left( s_0^{-\ell-1} H_1^{\ell-1} + t_0^{-\ell-1} K^{\ell -1} \right) \\
                  & \qquad + \Q_{\ell+1} \left( s_0^{\ell} H_1^{\ell+1} + t_0^{\ell} K^{\ell +1} \right) \bigg] \\
                  & + a^2 \bigg[ \Q_{\ell-1} \Q_\ell \left( u_0^{-\ell-1} h_1^{\ell-2} + v_0^{-\ell-1} h_0^{\ell-2} \right) \\
                  & \qquad + \Q_{\ell+1} \Q_{\ell+2} \left( u_0^{\ell} h_1^{\ell+2} + v_0^{\ell} h_0^{\ell+2} \right) \bigg],
\end{split}
\end{equation}
for the axial-led variables, and
\begin{equation}
\begin{split}
\label{eq:Zer_redef_K_2nd_order}
    K^{\ell} & = \left[ \frac{\la -2}{2r} - \frac{3 f_0}{r(3+\la r)} \right] \Phi_{(+)}^{\ell} + f_0 \dl_r \Phi_{(+)}^{\ell}  \\ 
                  & + \frac{\ii a m}{\rho} c_K^{\ell} \Phi_{(+)}^{\ell} + a^2 d_K^{\ell} \Phi_{(+)}^{\ell}  \\
                  & + a \bigg[ \Q_\ell \left( s_K^{-\ell-1} h_1^{\ell-1} + t_K^{-\ell-1} h_0^{\ell -1} \right)  \\
                  & \qquad + \Q_{\ell+1} \left( s_K^{\ell} h_1^{\ell+1} + t_K^{\ell} h_0^{\ell +1} \right) \bigg]  \\
                  & + a^2 \bigg[ \Q_{\ell-1} \Q_\ell \left( u_K^{-\ell-1} K^{\ell-2} + v_K^{-\ell-1} H_1^{\ell-2} \right)  \\
                  & \qquad + \Q_{\ell+1} \Q_{\ell+2} \left( u_K^{\ell} K^{\ell+2} + v_K^{\ell} H_1^{\ell+2} \right) \bigg],
\end{split}
\end{equation}
\begin{equation}
\begin{split}
\label{eq:Zer_redef_H1_2nd_order}
    H_1^{\ell} & = \left[ \frac{2r-3}{2rf_0} - \frac{3}{3+\la r}\right] \rho \Phi_{(+)}^{\ell} + \rho r \dl_r \Phi_{(+)}^{\ell}  \\
                  & + \frac{\ii a m}{\rho} c_H^{\ell} \Phi_{(+)}^{\ell}+ a^2 d_H^{\ell} \Phi_{(+)}^{\ell}  \\
                  & + a \bigg[ \Q_\ell \left( s_H^{-\ell-1} h_1^{\ell-1} + t_H^{-\ell-1} h_0^{\ell -1} \right)  \\
                  & \qquad + \Q_{\ell+1} \left( s_H^{\ell} h_1^{\ell+1} + t_H^{\ell} h_0^{\ell +1} \right) \bigg] \\
                  & + a^2 \bigg[ \Q_{\ell-1} \Q_\ell \left( u_H^{-\ell-1} K^{\ell-2} + v_H^{-\ell-1} H_1^{\ell-2} \right)  \\
                  & \qquad + \Q_{\ell+1} \Q_{\ell+2} \left( u_H^{\ell} K^{\ell+2} + v_H^{\ell} H_1^{\ell+2} \right) \bigg],
\end{split}
\end{equation}
for the polar led ones. The coefficients $s_i^\ell$ and $t_i^\ell$ can be further split as 
\begin{align}
& s_i^\ell = s_{0,i}^\ell + a s_{1,i}^\ell \qquad t_i^\ell = t_{0,i}^\ell + a t_{1,i}^\ell     
\end{align}
One has to choose the coefficients $c_i^\ell$, $d_i^\ell$, $s_i^\ell$, $t_i^\ell$, $u_i^\ell$ and $v_i^\ell$ (with $i=0,1,K,H$) such that if one inverts equations~\eqref{eq:RW_redef_h1_2nd_order}--\eqref{eq:Zer_redef_H1_2nd_order} to find a consistent expression for $\Phi_{(\pm)}^\ell$ and $\dl_r\Phi_{(\pm)}^\ell$. We can equate the second expression to the radial derivative of the first and, provided that the functions $h_0$, $h_1$, $K$ and $H_1$ are linearly independent, we can set to zero all the coefficients that multiply them. This requirement fixes uniquely the coefficients $c_i^\ell$ and $d_i^\ell$ and relates the coefficients $s_i^\ell$ to the $t_i^\ell$ and the $u_i^\ell$ to the $v_i^\ell$. This freedom of reparameterization will be exploited in the next steps to fully diagonalize the equations for the functions $\Phi_{(\pm)}^\ell$. 

By inserting the definitions~\eqref{eq:RW_redef_h1_2nd_order}--\eqref{eq:Zer_redef_H1_2nd_order} into the equations $Q_2=0$ and $Z_5=0$, one finds the following generalized Regge-Wheeler and Zerilli equations
\begin{equation}
\begin{split}
    \frac{\dd \Phi_{(\pm)}^{\ell}}{\dd r_{*}^2} & - \left( \rho^2 + V_{(\pm)}^{\ell} \right) \Phi_{(\pm)}^{\ell} \\
    & = a \left(A_{(\pm)}^\ell \Phi_{(\mp)}^{\ell+1} +A_{(\pm)}^{-\ell-1} \Phi_{(\mp)}^{\ell-1} \right) \\
    & + a^2 \left( B_{(\pm)}^\ell \Phi_{(\pm)}^{\ell+2} + B_{(\pm)}^{-\ell-1} \Phi_{(\pm)}^{\ell-2} \right) ,
\end{split}
\end{equation}
where the tortoise coordinate is $\dd r_* = \dd r /f_T$, being 
\begin{equation}\label{eq:tortoise_coordinate}
f_T = f_0 \left(1 + \frac{\ii a m f_1}{\rho} + a^2 f_2 \right),
\end{equation}
the functions $f_1$ and $f_2$ are left unspecified for now, as this can always be done by rescaling simultaneously the equation and $\Phi_{(\pm)}$. We aim to set $A_{(\pm)}^{\ell}=0$ and $B_{(\pm)}^{\ell}=0$. This result can be achieved by fixing completely the yet unspecified functions $t_i^\ell$ and $v_i^\ell$. The full list of the coefficients up to first-order in the spin is shown in appendix~\ref{app:coefficients}. The steps taken until now finally lead to the two completely decoupled equations
\begin{equation}\label{eq:Schrodinger}
    \frac{\dd \Phi_{(\pm)}^{\ell}}{\dd r_{*}^2} - \left( \rho^2 + V_{(\pm)}^{\ell} \right) \Phi_{(\pm)}^{\ell}  = 0 .
\end{equation}
The potentials are corrected as
\begin{align}\label{eq:potential_slow_rot}
    V_{(\pm)}^\ell = V_{(\pm),0}^\ell + \frac{\ii a m}{\rho}V_{(\pm),1}^\ell + a^2 V_{(\pm),2}^\ell,
\end{align}
If we do not specify the tortoise coordinate defined in equation~\eqref{eq:tortoise_coordinate}, the first and second-order potentials have the following form
\begin{align}
    V_{(\pm),1}^\ell = & \, \widebar{V}_{(\pm),1}^\ell  - \frac{1}{2}\frac{\dd^2 f_1}{\dd r_{*,0}^2} + 2 \left( \rho^2 + V_{(\pm),0}^\ell\right) f_1 \label{eq:potential_1} \\
    V_{(\pm),2}^\ell = & \, \widebar{V}_{(\pm),2}^\ell  - \frac{1}{2}\frac{\dd^2 f_2}{\dd r_{*,0}^2} + 2 \left( \rho^2 + V_{(\pm),0}^\ell\right) f_2 \notag \\
    &  - \frac{m^2}{\rho^2} \left[ \frac{3}{4}\left( \frac{\dd f_1}{\dd r_{*,0}} \right)^2 + 2 \widebar{V}_{(\pm),1}^\ell f_1 \right] \label{eq:potential_2}
\end{align}
where the corrections at first-order in the spin are
\begin{align}
    \widebar{V}_{(-),1}^\ell = & \,\frac{6f_0(7-6r)}{\ell(\ell+1)r^6} + \frac{2\rho^2}{r^3} , \label{eq:RW_1stcorrection} \\
    \widebar{V}_{(+),1}^\ell = & \, \frac{f_0}{(\la +2) r^7 (3+ r\la)^4} \bigg[6 \la ^4 (\la +4) r^6 \notag \\
    & \, +\la ^3 \left(-7 \la ^2-4 \la +96\right) r^5 \notag \\
    & \, -6 \la ^2 \left(5 \la ^2+8 \la -8\right) r^4 \notag \\ 
    & \, -3 \la 
   \left(29 \la ^2-84 \la +90\right) r^3 \notag \\
   & \, -18 \left(21 \la ^2-64 \la +18\right) r^2 \notag \\
   & \, -162 (6 \la -7) r-864\bigg] + \frac{2\rho^2}{r^3} \notag \\
   & \,+ 4 f_0 \frac{ (\la -2) \la ^2 r^3+9 \la ^2 r^2+15 \la  r+9}{(\la +2) r^3 (3 + r\la)^3}\rho ^2 , \label{eq:Zer_1stcorrection}
\end{align}
while second-order corrections are given in appendix~\ref{app:potentials}. We checked that if $f_1=0$ the first-order corrections to the two potentials are consistent with those already calculated in~\cite{Pani:2013pma}. With this calculation, we showed that despite rotation introduces coupling between different parities, it is nevertheless possible to diagonalize the system into two separate, axial-led and polar-led, modes. One can conjecture whether these modes are related to each other as in the non-rotating limit, and whether this behaviour persists at any order in the spin. While an answer to the second question is beyond the scope of this work, in the next section we show how the modes $\Phi_{(+)}$ and $\Phi_{(-)}$ are actually connected.

\section{Isospectrality of polar-led and axial-led equations}\label{sec:isospectrality}
The Regge-Wheeler and the Zerilli potentials are known to be {\it isospectral}. This means that, despite the functional form of the potentials in the two equations is drastically different, their spectrum of QNMs identically coincide. The motivation for this result was shown for the first time by Chandrasekhar~\cite{Chandrasekhar:1975nkd}, who realised that the two potentials can be generated by a {\it super-potential} $W_0^\ell$
\begin{equation}\label{eq:RWZ_generator}
    V_{(\pm),0}^\ell = \be_0^2 {W_0^\ell}^2 + \be_0 \frac{\dd W_0^\ell}{\dd r_{*,0} } + \ka_0 W_0^\ell ,
\end{equation}
with $\be_0 = \pm 3$, and the choice of the sign selects either the Regge-Wheeler or the Zerilli potential, $\ka_0 = \la(\la+2)$ and
\begin{equation}\label{eq:superpotential}
    W_0^\ell = \frac{f_0}{r(3+\la r)} .
\end{equation}
This simple-looking relation can be explained in the context of Darboux transformation, whose applications to BHPT have been extensively discussed in~\cite{Glampedakis:2017rar}. For the non-spinning case, the Darboux transformation is of the form~\cite{Chandrasekhar:1985kt}
\begin{equation}\label{eq:DT_RWZ}
    \Phi_{(\pm)}^\ell = \frac{\dd \Phi_{(\mp)}^\ell}{\dd r_{*,0}} + \left( \frac{\ka_0}{2\be_0} + \be_0 W_0^\ell \right) \Phi_{(\mp)}^\ell ,
\end{equation}
where a positive (negative) sign of $\be_0$ determines whether the transformation is from polar (axial) to axial (polar) functions.

We now want to analyse whether this structure is still existent for the slow-spinning potentials found in the previous section. We seek a {\it super-potential} $W^\ell$ that can generate the potentials $V_{(\pm)}^\ell$ through a generalized version of equation~\eqref{eq:RWZ_generator}
\begin{equation}\label{eq:RWZ_generator_generic}
    V_{(\pm)}^\ell = \be_0^2 {W^\ell}^2 + \be_0 \frac{\dd W^\ell}{\dd r_*} + \ka_0 W^\ell + \ka_0 \ka,
\end{equation}
where $\ka$ is a constant, the sign of $\be_0$ determines to retrieve the axial-led or polar-led potentials and we assume to truncate the calculation at second-order in the spin
\begin{align}
    W^\ell & = W_0^\ell + \frac{\ii a m}{\rho} W_1^\ell + a^2 W_2^\ell , \\
    \ka & = \frac{\ii a m}{\rho}  \ka_1 + a^2 \ka_2 .
\end{align}
From reference~\cite{Glampedakis:2017rar}, we know that the function $W^\ell$, if it exists, must satisfy the following constraints
\begin{align}
\label{eq:Darboux_deriv}
    \frac{\dd W^\ell}{\dd r_*} & = -\frac{1}{2} \left( V_{(-)}^\ell - V_{(+)}^\ell \right) \\
\label{eq:Darboux}
    W^\ell & = \frac{1}{2\left( V_{(-)}^\ell - V_{(+)}^\ell \right)} \frac{\dd}{\dd r_*} \left( V_{(-)}^\ell + V_{(+)}^\ell \right) .
\end{align}
By solving equation~\eqref{eq:Darboux_deriv} at first-order in the spin, we do it separately for each power of $\rho$. Hence, we find convenient to perform this further splitting
\begin{align}
    W_1^\ell & = \widetilde{W}_1^\ell + \widebar{W}_1^\ell \rho^2 , \\
    \ka_1 & =  \widetilde{\ka}_1 + \widebar{\ka}_1 \rho^2  .
\end{align}
For the terms proportional to $\rho$, we obtain
\begin{equation}
    \widebar{W}_1^\ell = \widebar{k}_1 - \frac{3+ 6 r \la + r^2 \la (\la-2)}{3 r^2 (\la+2) (3 + r \la)^2} , 
\end{equation}
where $\widebar{k}_1$ is an integration constant. By checking the self-consistency of the result with equation~\eqref{eq:Darboux}, we find that it is satisfied only for a specific choice of the tortoise coordinate~\eqref{eq:tortoise_coordinate}, which until now was left unspecified. Explicitly one has that the first-order function of the tortoise coordinate must assume the form
\begin{equation}\label{eq:Tortoise1_isospectrality}
f_1 = \frac{\ka_0 \widebar{\ka}_1 }{2} - \frac{1}{r^3} + \left( \frac{\ka_0}{2} + \be_0^2 W_0^\ell \right) \widebar{W}^\ell_1 - \frac{3}{2} \frac{\dd \widebar{W}^\ell_1}{\dd r_{*,0}} .
\end{equation}
In order to have consistency with the equations~\eqref{eq:Darboux_deriv}--\eqref{eq:Darboux} also for the terms proportional to $\rho^{-1}$, we find
\begin{align}
    & \frac{\widetilde{W}_1^\ell}{W_0^\ell} = \frac{r^2(3r-4)\la(\la+10) + 12r-36}{6r^4(\la+2)(3+r\la)} + 2 f_1 , \\
    & \widetilde{\ka}_1 = 0, \qquad \widebar{\ka}_1 = -\widebar{k}_1 = \frac{2}{9} .
\end{align}
The calculation at second-order in the spin proceeds analogously. Since it is longer, but essentially similar to that at first-order, we show it in appendix~\ref{app:generators}. Indeed, we found that in order to find a generator of the potentials, we need to fully specify the tortoise function $f_2$. Differently from the first-order case though, the consistency of the Darboux transformation does not fully specify all the integration constants.

With this calculation, we showed how one can transform the Regge-Wheeler equation to the Zerilli equation and vice-versa up to second-order in the spin, upon suitable choice of the tortoise coordinate. We conclude by showing explicitly the Darboux transformation between polar-led and axial-led quantities
\begin{equation}\label{eq:DT_RWZ_2nd_order}
    \Phi_{(\pm)}^\ell = \frac{\dd \Phi_{(\mp)}^\ell}{\dd r_*} + \left( \frac{\ka_0}{2\be_0} + \be_0 W^\ell \right) \Phi_{(\mp)}^\ell ,
\end{equation}
The existence of this transformation implies that the spectrum of the two potentials is exactly equivalent, since it does not modify the boundary conditions, as it happens in the non-rotating case~\cite{Glampedakis:2017rar}. The only difference is that the Darboux relation is manifest only for one specific choice of the tortoise coordinate. We guess that one can anyway find a more general transformation that links the two equations if the tortoise coordinate is left unspecified.  In any case, we want to stress that there is nothing special with truncating the expansion at second-order in the spin, and we expect this trend to continue even at higher orders. We will now turn the focus to the relation between the Regge-Wheeler/Zerilli equations and the Teukolsky equation in the slow rotation limit.

\section{Isospectrality with slow-spinning Teukolsky equation}\label{sec:iso_teuk}
\subsection{The Teukolsky equation}
BHPT for Kerr BHs is most commonly studied within the Teukolsky formalism~\cite{Teukolsky:1973ha}. This approach works by projecting the spacetime onto the null Newman-Penrose tetrad, and by analysing the geometrical relations between some components of the projected Weyl tensor, which is the trace-free Riemann tensor. All the information of the projected Weyl tensor is contained in five complex scalars. By considering the perturbation of these Weyl scalars, only two of them cannot be set to zero with an infinitesimal gauge transformation. We commonly refer to them as $\Psi^{(0)}$ and $\Psi^{(4)}$, which can be treated as linear perturbations quantities, since their background vanishes for the Kerr solution. The Teukolsky equations are linear second-order differential equations for $\Psi^{(0,4)}$, which can be considered as perturbations of internal spin $s=\pm2$, but it is valid also for $s=0$ (scalar perturbations), $s=\pm1/2$ (spinor perturbations) and $s=\pm1$ (electromagnetic perturbations). Its entire derivation can be found in detail in~\cite{Chandrasekhar:1985kt}.
Here, we report the general form of the Teukolsky master equation
\begin{equation}\label{eq:teukolsky}
\begin{split}
    & \left[\frac{\left(r^2 + a^2\right)^2}{\De} - a^2 \sst \right] \dl^2_t \psi^{(s)} 
    + \frac{2ar}{\De} \dl_t \dl_{\phi}\psi^{(s)} \\
    & + \left[\frac{a^2}{\De} - \frac{1}{\sst} \right] \dl^2_{\phi} \psi^{(s)} - \De^{-s} \partial_r \left( \De^{s+1} \partial_r \psi^{(s)} \right) \\
    & - \frac{1}{\sin{\th}} \dl_{\th} \left( \sin{\th} \dl_{\th} \psi^{(s)} \right) + (s^2 \cot{\th}^2 -s ) \psi^{(s)} \\
    & - 2s \left[\frac{a \left(2r-1 \right)}{2\De}+\ii \frac{\cos{\th}}{\sst} \right] \dl_{\phi} \psi^{(s)}
    \\
    & -2s \left[ \frac{(r^2-a^2)}{2\De} - r - \ii a \cos{\th} \right] \dl_t \psi^{(s)} 
     = 0 .
\end{split}
\end{equation}
For the analysis pursued in this paper, we are only interested in tensor perturbations ($s=\pm2$), for which the perturbation function $\psi^{(s)}$ can be linked to first-order perturbations of the Weyl scalars as
\begin{equation}\label{eq:psiTeukolsky}
    \psi^{(2)} = \Psi^{(0)}, \qquad \psi^{(-2)} = (r - \ii a \cos\th)^4 \Psi^{(4)}.
\end{equation}
In the Teukolsky equation one can decouple the radial and the angular part, by choosing the following mode decomposition
\begin{equation}\label{eq:TeukolskyAnsatz}
    \psi^{(s)} = R^{(s)}_{\ell m}(r) S^{(s)}_{\ell m}(\th) \, \ee^{\rho t + \ii m \cf } .
\end{equation}
For the angular part one gets
\begin{equation}\label{eq:SpheroidalHarmonics}
\begin{split}
    \frac{1}{\sin\th} \dl_\th\left( \sin\th \dl_\th S_{\ell m}^{(s)} \right)  + \bigg[ 2 \ii s a \rho \cos\th-a^2 \rho^2 \cct & \\
    - \frac{\left(m + s \cos\th\right)^2}{\sst} + s + \la_{\ell m}^{(s)} \bigg] S_{\ell m}^{(s)} = 0 & ,
\end{split}
\end{equation}
where the functions $S_{\ell m}^{(s)}$ are known as {\it spin-weighted spheroidal harmonics} and $\la_{\ell m}^{(s)}$ is the separation constant and it depends on the frequency $\rho$. We find it more convenient to work with the reduced constant $\lambdabar_{\ell m}^{(s)} = \la_{\ell m}^{(s)} + (s + |s|)$. The constants $\lambdabar_{\ell m}^{(s)}$ are only known numerically, but we report here their value up to second-order in the spin  $\lambdabar^{(s)}_{\ell m} \simeq \la + \ii a m \rho \la^{(s)}_1 + a^2\rho^2\la^{(s)}_2$, where~\cite{Berti:2005gp}
\begin{align}
    \la_1^{(s)} = -\frac{2s^2}{\la+2} , \qquad
    \la_2^{(s)} = \frac{1 + \mathcal{P}^{(s)}_{\ell} -\mathcal{P}^{(s)}_{\ell+1}}{\la+2} ,
\end{align}
where we defined
\begin{equation}
    \mathcal{P}^{(s)}_{\ell} = 2\Q_\ell^2 \frac{(\ell^2 - s^2)^2}{\ell^3} .
\end{equation}
Finally, the radial part of the Teukolsky equation must satisfy the following equation
\begin{align}\label{eq:Teukolsky_Radial}
    \De^{-s} \dl_r & \left( \De^{s+1} \dl_r \right) R_{\ell m}^{(s)} \notag \\
    + & \bigg[ \frac{K^2 - \ii s(2r - 1)K}{\De} - a^2 \om^2 \notag \\
    + & 2 m a \om +  4\ii s \om r  - \la_{\ell m}^{(s)}\bigg] R_{\ell m}^{(s)} = 0,
\end{align}
with $K = \left(r^2 + a^2 \right) \om - a m $.

\subsection{The isospectrality relation}\label{subsec:isospectrality_teukolsky}
In order to show the isospectrality between metric and Weyl scalar perturbations up to second-order in the spin, we borrow the notation to write the equations introduced by Chandrasekhar in~\cite{Chandrasekhar:1985kt}. First of all, we introduce the operators
\begin{equation}
    \La_\pm = \frac{\dd}{\dd r_*} \pm \rho .
\end{equation}
Then, after redefining the radial functions as
\begin{equation}\label{eq:TekolskyY}
    U^{(s)}(r) = \frac{\De^{(s\pm2)/2}}{r^3} R^{(s)}(r),
\end{equation}
one can compactly write the Teukolsky equation for the perturbation variable $U$\footnote{The two $\pm s$ polarizations defined in this way satisfy complex conjugate equations~\cite{Chandrasekhar:1985kt}. Moreover, the separation constants $\la_i^{(s)}$ are identical for both choices of the spin. From now on, we will focus on $s=2$ only and drop the index $s$ everywhere.}
\begin{equation}\label{eq:Teuk_Chandra}
    \La_+ \La_- U + P \La_- U - Q U = 0 ,
\end{equation}
where
\begin{equation}
    P = \frac{\dd}{\dd r_*} \log\left( \frac{f_P^2}{f_T^2}\right) ,
\end{equation}
where we remind that $f_T$ is the tortoise function defined in equation~\eqref{eq:tortoise_coordinate} and $f_P$ and $Q$ are functions that must be specified such that they match the Teukolsky equation. We found that a convenient definition is
\begin{equation}\label{eq:Pteuk}
    f_P = \frac{r^2}{2} \left[ 1 + \frac{f_T}{f_0} + a^2 \left(\frac{m^2}{\rho^2} \frac{f_1^2}{4} - \frac{1}{r^2 f_0} \right)\right] .
\end{equation}
With this choice, the effective potential takes schematically the form
\begin{equation}\label{eq:VT_format}
    Q = Q_{0} + \frac{\ii a m}{\rho} Q_{1} + a^2 Q_{2} ,
\end{equation}
where we labelled as $Q_0$ the effective potential of the Bardeen-Press equation, which we know from the relations obtained by Chandrasekhar~\cite{Chandrasekhar:1985kt} that it can be related with the super-potential $W_0$ defined in equation~\eqref{eq:superpotential} as
\begin{equation}\label{eq:BardeenPressPotential}
    Q_{0} = \frac{f_0^2}{r^4 W_0} ,
\end{equation}
and we furthermore split first and second-order corrections by powers of $m$ and $\rho$ as
\begin{align}
\label{eq:VT1_format}
    Q_{1} & = \sum_{i=0}^2 Q_{1,i} \rho^i\\
    Q_{2} & = \sum_{i=0}^2 Q_{2,i} \rho^i + \frac{m^2}{\rho^2} \sum_{i=1}^2 \widebar{Q}_{2,i} \rho^i
\label{eq:VT2_format},
\end{align}
with $Q_{i,j}$ and $\widebar{Q}_{i,j}$ functions of $r$ and $\ell$. The choice for $f_P$ given in equation~\eqref{eq:Pteuk} is such that the first-order expressions for $Q_{1}$ take the rather simple form form
\begin{align}
    Q_{1,0} & = 2 Q_0 f_1 \\
    Q_{1,1} & = \frac{\dd f_1}{\dd r_{*,0}} - 2\frac{3-2r}{r^2}f_1 + 2 \frac{1-2r}{r^4} \\
    Q_{1,2} & = \frac{2}{r^3} + 2  f_1 + f_0 \frac{\la_1}{r^2},
\end{align}
whereas the expression of $Q_2$ is
\begin{align}
    Q_{2,0} & = \frac{3-r(\la+6)}{r^5} + 2 Q_0 f_2 , \\
    Q_{2,1} & = \frac{\dd f_2}{\dd r_{*,0}} - 2 \frac{3-2r}{r^2}f_2 + \frac{(11-6r)r-2}{f_0 r^5} , \\
    Q_{2,2} & = -\frac{r^2 + 1}{r^4 f_0} + 2 f_2 + f_0 \frac{\la_2}{r^2} , \\
    \widebar{Q}_{2,1} & = \left[ Q_{1,1} - \frac{3}{r^2}\left(2 \frac{1-2r}{r^2} + (2r-3)f_1 \right)\right] f_1 , \\
    \widebar{Q}_{2,2} & = - 2 \left(Q_{1,2} - 2f_1\right) f_1 .
\end{align}

In the limit $a\to 0$, the Teukolsky equation reduces to the Bardeen-Press equation.
In~\cite{Chandrasekhar:1985kt}, Chandrasekhar found an elegant way, which he labelled {\it transformation theory}, to transform it to the Regge-Wheeler or to the Zerilli equation, which in the notation used in this section appear as
\begin{equation}\label{eq:RWZ_Chandra}
    \La_+ \La_- \Phi_{(\pm)} - V_{(\pm)} \Phi_{(\pm)} = 0 ,
\end{equation} 
and back. The transformation found by Chandrasekhar is more general than the Darboux transformation used in section~\ref{sec:isospectrality}, and falls within the class of generalized Darboux transformations~\cite{Glampedakis:2017rar}.
It is remarkable that the {\it transformation theory} embeds naturally the isospectrality relation between the Regge-Wheeler and the Zerilli equation, as we made manifest by explicitly showing the relation between the Bardeen-Press potential $Q_0$ and the super-potential $W_0$ in equation~\eqref{eq:BardeenPressPotential}. We now want to see if the slowly-rotating Regge-Wheeler and Zerilli equations can be linked to the slowly-rotating expansion of the Teukolsky equation in a similar manner.

Let us revise how to obtain the transformation proposed by Chandrasekhar, slightly generalizing it. The first step is to assume that field $U$ transforms as\footnote{From now on, we drop the label $\pm$ to the Regge-Wheeler and Zerilli functions, as each transformation must be intended for the two fields separately, but has the same functional form.}
\begin{equation}\label{eq:transformation_U}
    U = \f \La_+^2 \Phi + \left( T - 2\rho \f \right) \La_+ \Phi,
\end{equation}
where $\f$ and $T$ are the functions that we want to find. Alternatively, by using the fact that $\Phi$ satisfies equation~\eqref{eq:RWZ_Chandra}, one can rewrite the transformation as
\begin{equation}\label{eq:TeukoToRWZ}
    U = \f V \Phi + T \La_+ \Phi.
\end{equation}
The application of the operator $\La_-$ to both sides of the equations yields
\begin{equation}\label{eq:transformation_LaU}
    \La_- U = - \frac{f_T^2}{f_P^2} \be \Phi + R \La_+ \Phi
\end{equation}
where we introduced the functions $\be$ and $R$ defined from the two relations
\begin{align}
\label{eq:rel1}
   \Rel_{1} & \equiv \frac{f_T^2}{f_P^2} \be + \frac{\dd}{\dd r_*} \left( \f V \right) + \left( T - 2\rho\f \right) V = 0 , \\
\label{eq:rel2}
   \Rel_2 & \equiv R - \f V - \frac{\dd T}{\dd r_*} = 0.
\end{align}
Now, we need to require that the function $Y$ satisfies the equation~\eqref{eq:Teuk_Chandra}, which leads to two additional relations for the free functions that we defined
\begin{align}
\label{eq:rel3}
   \Rel_3 & \equiv R V - \frac{f_T^2}{f_P^2} \frac{\dd \be}{\dd r_*} - \f Q V = 0 , \\ 
\label{eq:rel4}
   \Rel_4 & \equiv \frac{\dd R}{\dd r_*}  + R P - \left(Q T - 2\rho R\right) - \frac{f_T^2}{f_P^2}\be = 0 .
\end{align}
Finally, it is worth noticing that one can combine the equations~\eqref{eq:rel1}--\eqref{eq:rel4} into one single integral relation
\begin{equation}
\label{eq:rel5}
   \Rel_5 \equiv \frac{f_P^2}{f_T^2} R V \f + \be T = K = \const.
\end{equation}
This last relation assures that, if a transformation is found, also the inverse transformation exists
\begin{align}\label{eq:RWZtoTeuk}
    K \Phi & = \frac{f_P^2}{f_T^2}  \left( R U - T \La_- U \right) , \\
    K \La_+ \Phi & = \be U + \frac{f_P^2}{f_T^2} V \f \La_- U .
\end{align}
Now we assume that the unknown functions $\f$, $R$, $T$ and $\be$ as well as the constant $K$ can be expanded in powers of $a$, $m$ and $\rho$ in the same form of $Q$ given in equations~\eqref{eq:VT_format},~\eqref{eq:VT1_format} and~\eqref{eq:VT2_format}. In the limit $a\to 0$, they must reproduce the same value found by Chandrasekhar for the generalized Darboux transformation between the Bardeen-Press and the Regge-Wheeler/Zerilli equations
\begin{align}
    & \f_0 = 1, \qquad \be_0 = \pm 3, \qquad R_0 = Q_{0}, \\
    & K_0 = \ka_0 , \quad \widetilde{T}_0 = \frac{1}{\be_0}\left( \ka_0 - \frac{V_{0}}{W_0} \right),
\end{align}
and $T_0 = \widetilde{T}_0 + 2\rho$.
For the next orders in the spin we solve the equations $\Rel_1$, $\Rel_2$, $\Rel_4$ and $\Rel_5$, assuming that they vanish independently for each power in $\rho$ and $m$. We were able to find a solution to the problem, as schematically shown below.

At first-order in the spin the functions to be specified are $\f_{1,i}$, $\be_{1,i}$, $T_{1,i}$, $\widetilde{R}_{1,i} = R_{1,i} - Q_{1,i}$ and $K_{1,i}$ with $i = [ 0,1,2]$. We also found convenient to rewrite $V_1 = V_{1,0} + V_{1,2} \rho^2$. We find terms proportional to $\rho^3$ only in $\Rel_1$, $\Rel_4$ and $\Rel_5$. By making them vanish, we find that
\begin{equation}
    \f_{1,2} = \be_{1,2} = \widetilde{R}_{1,2} = 0.
\end{equation}
We can manipulate the terms proportional to $\rho^2$ into three algebraic relations and one first-order differential equation. Their solution reads
\begin{align}
    & 2\f_{1,1} = T_{1,2} + \widetilde{T}_0 \frac{V_{1,2}}{V_0} + \frac{1}{V_0}\frac{\dd V_{1,2}}{\dd r_{*,0}}, \\
    & 2\be_{1,1} = K_{1,2} - \frac{1}{W_0} \left( \frac{Q_{1,2} V_0}{Q_0} + V_{1,2} \right) - \be_0 T_{1,2}, \\
    & 2\widetilde{R}_{1,1} = Q_0 T_{1,2} + \left( \widetilde{T}_0 + 2\frac{3-2r}{r^2} \right) Q_{1,2} - \frac{\dd Q_{1,2}}{\dd r_{*,0}} ,\\
    & T_{1,2} = -\frac{\la_1}{r} - (3+\be_0) \left(\widebar{W}_1 -\widebar{k}_1 \right) + t_2,
\end{align}
where $t_2$ is an integration constant.
Analogously, the terms proportional to $\rho$ can be written as three algebraic relations and one first-order differential equation, whose solution is
\begin{align}
    2\f_{1,0} & = T_{1,1} - \widetilde{T}_0 \f_{1,1} \notag \\
    & + \frac{W_0 Q_0}{V_0} \left( \be_{1,1} - \be_0 \f_{1,1} \right) + \frac{\dd \f_{1,1}}{\dd r_{*,0}} \\
    2\be_{1,0} & = K_{1,1} + \frac{V_0}{W_0} \left( \frac{\be_{1,1}}{\be_0} - \f_{1,1} - \frac{R_{1,1}}{Q_0} \right) \notag \\
    & - \be_0 T_{1,1} - \frac{\ka_0\be_{1,1}}{\be_0} \\
    2\widetilde{R}_{1,0} & = Q_0 \left(T_{1,1} + \be_{1,1}W_0 \right) + \widetilde{T}_0 Q_{1,1} \notag \\
    & + 2\frac{3-2r}{r^2} R_{1,1} - \frac{\dd R_{1,1}}{\dd r_{*,0}} \\
    2 T_{1,1} & = T_{1,2}\widetilde{T}_0-V_{1,2} - Q_{1,2} - 2f_1 \notag \\
    & \, +2f_0\frac{\la_1}{r^2} +\frac{4(r+1)}{r^3} + \ka_0\widebar{\ka}_1 + t_1,
\end{align}
where $t_1$ is another integration constant. Finally, from the last group of equations we obtain
\begin{align}
    T_{1,0} & = \frac{\dd f_1}{\dd r_{*,0}} + \frac{1}{Q_0}\frac{\dd R_{1,0}}{\dd r_{0,*}} - \be_{1,0} W_0  \notag \\
    & - 2 \frac{3-2r}{r^2}\frac{R_{1,0}}{Q_0} + \left( \widetilde{T}_0 + 2\be_0 W_0 \right) f_1 \\
    K_{1,2} & = 2\be_0 t_2 -4\la , \qquad K_{1,1} = 2\be_0 t_1 + \ka_0 t_2 , \\
    K_{1,0} & = \ka_0 t_1 .
\end{align}
We checked that these relations can be found only for the same choice of the tortoise coordinate $f_1$ as given in equation~\eqref{eq:Tortoise1_isospectrality}. This shows that the deep intimacy between metric perturbations and the Teukolsky formalism is maintained at first-order in the spin. We argue that a different choice for the tortoise coordinate can still admit a generalized Darboux transformation between the Regge-Wheeler, the Zerilli and the Teukolsky equations at first-order in the spin, but not with a compact form polynomial in $\rho$. It is worth noting that the integrating constants $t_1$ and $t_2$ are left completely unspecified from the calculation.

The calculation at second-order is complicated by the fact of having more free functions, but it can be performed in the same fashion: treating each term with a different power in $\rho$ and in $m$ as independent and setting it to zero. In this way, we were able to find a consistent solution, provided that the tortoise coordinate is the same selected by the isospectrality analysis of the Regge-Wheeler and Zerilli potentials described in section~\ref{sec:isospectrality}. The whole generalized Darboux transformation is presented in appendix~\ref{app:transformation}.

\section{Discussion}\label{sec:discussion}

\subsection{The metric reconstruction}

Throughout the calculations performed in this paper, we managed to find a set of analytic transformations that bring slowly rotating Regge-Wheeler and Zerilli equations to the spin-$2$, slow spin limit of the Teukolsky equations and vice versa. The second direction is of particular interest for various fields of gravitational physics, because it allows one to obtain the perturbations of the metric $h_{ab}$ once the gauge invariant Weyl scalars $\Psi^0$ and $\Psi^4$ are known. We summarize the main steps of this transformation, directly linking to the necessary equations provided in the paper
\begin{itemize}
    \item The Weyl scalars $\Psi^0$ and $\Psi^4$ must be decomposed according to equations~\eqref{eq:psiTeukolsky}--\eqref{eq:TeukolskyAnsatz}. The radial functions $R_{\ell m}^{(\pm2)}(r)$ can then be transformed to the variables $U_{\ell m}^{(\pm2)}(r)$ as specified in equation~\eqref{eq:TekolskyY}.
    \item The relation that tells how to transform the Teukolsky variable to either the Regge-Wheeler and the Zerilli variables is given in equation~\eqref{eq:RWZtoTeuk}. This relation depends on the quantities $R$, $T$, and $K$, which have been provided up to second-order in the spin. It is worth noting that these quantities depend on the parameter $\be_0=\pm3$, for which the sign choice tells how $\Phi_{(\pm)}^{\ell m}(r)$ are given in terms of $U_{\ell m}^{(2)}(r)$ and its radial derivative (up to an integrating constant). The transformations for $U^{(-2)}_{\ell m}$ can be found analogously by following the calculations provided in this paper.
    \item Once one knows $\Phi_{(\pm)}^{\ell m}(r)$, the functions $h_0^{\ell m}(r)$, $h_1^{\ell m}(r)$, $H_1^{\ell m}(r)$ and $K^{\ell m}(r)$ can be constructed from equations~\eqref{eq:RW_redef_h1_2nd_order}--\eqref{eq:Zer_redef_H1_2nd_order}. Indeed, by repeated use of these equations, and by keeping terms up to second-order in $a$, one can write
\begin{widetext}
    \begin{equation}
    \begin{split}
        h_{i,(\pm)}^{\ell} = & \, c_{i,(\pm)}^{\ell} \Phi_{(\pm)}^\ell + d_{i,(\pm)}^{\ell} \dl_r \Phi_{(\pm)}^\ell + a \left[ \Q_\ell \left( s_{i,(\pm)}^{-\ell-1}\Phi_{(\mp)}^{\ell - 1} + t_{i,(\pm)}^{-\ell-1}\dl_r\Phi_{(\mp)}^{\ell - 1} \right) + \Q_{\ell+1} \left( s_{i,(\pm)}^{\ell}\Phi_{(\mp)}^{\ell + 1} + t_{i,(\pm)}^{\ell+1}\dl_r\Phi_{(\mp)}^{\ell + 1} \right) \right] \\
        & \, + a^2 \left[ \Q_{\ell-1}\Q_{\ell} \left( u_{i,(\pm)}^{-\ell-1} \Phi_{(\pm)}^{\ell-2} + v_{i,(\pm)}^{-\ell-1} \dl_r \Phi_{(\pm)}^{\ell-2} \right) + \Q_{\ell+1}\Q_{\ell+2} \left( u_{i,(\pm)}^{\ell} \Phi_{(\pm)}^{\ell+2} + v_{i,(\pm)}^{\ell+1} \dl_r \Phi_{(\pm)}^{\ell+2} \right) \right],
    \end{split}
    \end{equation}
\end{widetext}
    where $h_{0,(-)}^\ell = h_0^\ell$, $h_{1,(-)}^\ell = h_1^\ell$, $h_{K,(+)}^\ell = K^\ell$ and $h_{1,(+)}^\ell = H_1^\ell$, and the coefficients $c_{i,(\pm)}^\ell$, $d_{i,(\pm)}^\ell$, $s_{i,(\pm)}^\ell$, $t_{i,(\pm)}^\ell$, $u_{i,(\pm)}^\ell$ and $v_{i,(\pm)}^\ell$ can be related to the coefficients $c_i^\ell$, $d_i^\ell$, $s_i^\ell$, $t_i^\ell$, $u_i^\ell$ and $v_i^\ell$ introduced in section~\ref{subsec:RWZequations}.
    Finally, the remaining functions $H_0^{\ell}(r)$ and $H_2^{\ell}(r)$ are determined by the constraints set by the polar-led equations and discussed in section~\ref{subsec:RWZequations}.
\end{itemize}
All of these steps completely determine how to construct, in vacuum, the first-order perturbation metric $h_{ab}$ in the Regge-Wheeler gauge, up to second-order in the spin starting the from the knowledge of the Weyl scalars.

\subsection{The generator of the slowly rotating Regge-Wheeler and Zerilli effective potentials}

Another very remarkable result of this paper is that we could find the generalization of Chandrasekhar {\it super-potential} up to second-order in the spin. 
This allows one to have a handy and compact formula to generate the first-order to the Regge-Wheeler and Zerilli equations~\eqref{eq:RW_1stcorrection}--\eqref{eq:Zer_1stcorrection}
\begin{equation}
\begin{split}
    V^\ell_{(\pm),1} = \left( \ka_0 + 2\be_0^2 W_0^\ell \right) W_1^\ell + \ka_0 \ka_1 & \\
    + \be_0 f_0 \left[\frac{\dd}{\dd r} W_1^\ell + f_1 \frac{\dd W_0^\ell}{\dd r}\right] & ,
\end{split}
\end{equation}
and second-order corrections (see appendix~\ref{app:potentials})
\begin{equation}
\begin{split}
    V^\ell_{(\pm),2} = \left( \ka_0 + 2\be_0^2 W_0^\ell \right) W_2^\ell + \ka_0 \ka_2 & \\
    + \be_0 f_0 \left[\frac{\dd}{\dd r} W_2^\ell + f_2 \frac{\dd W_0^\ell}{\dd r}\right] & \\
    - \be_0 \frac{m^2}{\rho^2} \left[ \be_0 {W_1^\ell}^2 + f_1 f_0 \frac{\dd W_1^\ell}{\dd r} \right] & .
\end{split}
\end{equation}
The explicit expressions of $W_0^\ell$, $W_1^\ell$, $\ka_0$, $\ka_1$ and $f_1$ is given in section~\ref{sec:isospectrality}, whereas expressions for $W_2^\ell$, $\ka_2$ and $f_2$ can be found in appendix~\ref{app:generators}. 

We want to stress that the function $W^\ell$, and the selection of the tortoise coordinate $f_T$ are naturally consistent with the procedure that links the slowly rotating Teukolsky potential with the Regge-Wheeler and the Zerilli potentials. This procedure also leaves unspecified some integration constants. We argue that this result can be interpreted as the freedom of scaling the functions $U_{\ell m}^{(\pm2)}(r)$ through the Teukolsky-Starobinsky identities~\cite{Teukolsky:1974yv,1974JETP...38....1S}.

\subsection{The Kerr metric perturbation conjecture}

The deep link between metric perturbations and curvature perturbations revealed by Chandrasekhar's {\it transformation theory} solidly holds up to second-order in the spin. For this reason we conjecture the existence of a yet to be discovered couple of equations that generalize the Regge-Wheeler and the Zerilli equations for any value in the spin. These equations would describe the perturbations of a Kerr metric, rather than the perturbations of its curvature as it happens in the Teukolsky equations. Past investigations failed in finding these equations, but we hope that the results of this paper can fuel a renovated research in this direction. The possible discovery of such equations would give insights about the isospectrality of modes of definite parity for fully spinning metric (see {\it e.g.}~the appendix of~\cite{Nichols:2012jn} for a discussion on reconstructing parity definite metric perturbations on a Kerr background)

We want to remark that we compared the slowly rotating Regge-Wheeler and Zerilli potentials with those obtained by Chandrasekhar and Detweiler in~\cite{Chandrasekhar:1976zz}. The Chandrasekhar-Detweiler potentials are obtained by performing a generalized Darboux transformation to the Teukolsky equation that brings it in a Schr\"odinger-like form. Among these four potentials, two reduce to the Regge-Wheeler potential in the non-rotating limit, while the other two reduce to the Zerilli potential. Moreover, it was shown numerically that the lowest-order quasi-normal modes of one of the Chandrasekhar-Deteweiler potentials that reduces to the Regge-Wheler equation, agree with those of computed from the Teukolsky equation until second-order in the spin~\cite{Hatsuda:2020egs}. Unfortunately, the shape of the first-order correction in the spin to the Chandrasekhar-Detweiler potentials is always different from the corrections found in this paper~\eqref{eq:RW_1stcorrection}--\eqref{eq:Zer_1stcorrection}. We also tried to see if transforming the equations into a different gauge or by using a different tortoise coordinate could bring the potentials in the same form, but it was not possible. We argue that the two classes of potentials must then be linked by a generalized Darboux transformation, even if we were not able to prove it.

\section{Conclusions}\label{sec:conclusions}

In this paper we provided a complete formalism to study vacuum BHPT in the regime of slow rotation up to second-order in the spin. We described how to perturb a slowly-spinning Kerr metric in the Regge-Wheeler gauge in the frequency domain, and the prescription to decouple the radial and the angular contribution to the equations. In this way we obtained seven {\it polar-led} equations and the three {\it axial-led} equations where each mode of angular momentum $\ell$ couples starting from the first-order in the spin to modes of different parity and angular momentum $\ell\pm1$ and starting from second-order in the spin to modes of same parity and $\ell\pm2$. We showed that a suitable redefinition of the variables allows one to decouple completely modes of different angular momentum and parity, leading to two diagonalized second-order differential equations that generalize the Regge-Wheeler and the Zerilli equation up to second-order in the spin. 

We then proved that these two potentials are not independent from each other, as a transformation that brings one to the other and vice versa was found. The existence of this transformation ensures the {\it isospectrality} of the two potentials, as well as the existence of a function that generates them. This generating function, also known as {\it super-potential} can be understood by comparing the metric perturbation equations to the Teukolsky equations. Indeed, we found that a third transformation that links Teukolsky, Regge-Wheeler and Zerilli equations still exists at second-order in the spin. We discussed how the existence of this transformation naturally embeds a procedure of metric reconstruction.

The approach taken in the paper does not seem to single out a reason why these results should not hold at any higher order in the spin expansion. The only impediment is the increasing difficulty in the calculations, especially for the computation of the integrals that appear in the process of decoupling the equations of motion, as those listed in appendix~\ref{app:integrals}. For this reason it would be revolutionary to find a prescription to study gravitational perturbations of a Kerr metric for any spin. Moreover, we expect that the covariant and gauge-independent formalism developed in~\cite{Lenzi:2021wpc,Lenzi:2021njy} should also hold at any order in the spin.

The results of this paper are very relevant for different fields of application. The difficulty of studying quasi-normal modes of rotating solutions in alternative theories of gravity requires some sort of simplification of the problem. One possibility often encountered is to evaluate them perturbatively in the spin, as it was done in~\cite{Cano:2021myl,Pierini:2021jxd,Wagle:2021tam,Srivastava:2021imr,Pierini:2022eim}. It would be interesting to see if, at least in the small coupling limit, one would still get a slowly rotating Regge-Wheeler and a Zerilli equation plus a correction due to the modifications of GR. By writing the problem in this form, one can use more accurate methods for the computation of the quasi-normal modes, rather than the direct integration usually employed in these cases, {\it e.g.}, the continued fraction method~\cite{Leaver:1985ax}.

In the context of quasi-normal modes computation beyond GR, there were recent progress in developing a generalized Teukolsky equation for any metric that is modified from the Kerr metric by a small parameter~\cite{Li:2022pcy,Hussain:2022ins,Cano:2023tmv}. Such a derivation requires the knowledge of metric perturbations, which appear as a ``source'' term for the perturbations of the Weyl scalars. Despite the procedure outlined in this paper is only perturbative in the spin, we stress that all the terms that require the metric reconstruction are multiplied by the small perturbative parameter of the theory. In an analogue case for scalar field perturbations on top of a non-Kerr BH, it was shown that by treating this term in a small-spin expansion does not affect strongly the computation of the quasi-normal modes~\cite{Ghosh:2023etd}.

Finally, we believe that it would be interesting to generalize the metric reconstruction procedure to the case where a point particle source is present in the spacetime, along the lines of~\cite{Lousto:2005xu}. This prescription would be useful in the context of self-force calculations, as it was noted that second-order contributions are necessary to compute accurate waveforms of extreme mass ratio inspirals~\cite{Wardell:2021fyy}.

\acknowledgements
I am indebted to Stas Babak, Enrico Barausse, Emanuele Berti, \'Eric Chassande-Mottin, Kostas Glampedakis, Leonardo Gualtieri, Luca Santoni, Sebastian V\"olkel and Niels Warburton for useful discussions, fruitful comments on the interpretation of the results, and the precious feedback on the manuscript.

\appendix

\section{Useful integrals}\label{app:integrals}
According to the author's knowledge, the integrals appeared in the decoupling of the equations in section~\ref{sec:angular_separation} cannot be found explicitly computed in the existing literature. For this reason we decide to report them here for the interest reader. All the expressions are obtained using simplifications due to the completeness relation of spherical harmonics~\eqref{eq:completeness}, as well as the spherical harmonics relation~\eqref{eq:spherical_hoarmonics}. By recursively applying the mixing of the spherical harmonics and their derivatives with the trigonometric functions~\eqref{eq:cosY}--\eqref{eq:sinY}, we obtain the following expression:
\begin{widetext}
\begin{align}
    \C_1 f_{\ell} & \equiv f_{\ell'} \int \dd\Om \, \cos\th \, Y^\ell Y^{*\ell'} = f_{\ell -1} \mathcal{Q}_{\ell }+f_{\ell +1} \mathcal{Q}_{\ell +1} , \\
    \C_2 f_{\ell} & \equiv f_{\ell'} \int \dd\Om \, \cos^2\!\th \, Y^\ell Y^{*\ell'} = f_{\ell -2} \mathcal{Q}_{\ell -1} \mathcal{Q}_{\ell }+f_{\ell } \left(\mathcal{Q}_{\ell }^2+\mathcal{Q}_{\ell +1}^2\right)+f_{\ell +2} \mathcal{Q}_{\ell +1}
   \mathcal{Q}_{\ell +2} , \\
    \Sc_{1} f_{\ell} & \equiv f_{\ell'} \int \dd\Om \, \sin\th \, Y_{,\th}^\ell Y^{*\ell'} = (\ell -1) f_{\ell -1} \mathcal{Q}_{\ell }-(\ell +2) f_{\ell +1} \mathcal{Q}_{\ell +1} , \\
    \Sc_{2} f_{\ell} & \equiv f_{\ell'} \int \dd\Om \, \sin\th \cos\th \, Y_{,\th}^\ell Y^{*\ell'} = (\ell -2) f_{\ell -2} \mathcal{Q}_{\ell -1} \mathcal{Q}_{\ell }+f_{\ell } \left(\ell  \mathcal{Q}_{\ell +1}^2-(\ell +1) \mathcal{Q}_{\ell }^2\right)-(\ell +3) f_{\ell +2}
   \mathcal{Q}_{\ell +1} \mathcal{Q}_{\ell +2} , \\
    \widebar{\Sc}_{1} f_{\ell} & \equiv f_{\ell'} \int \dd\Om \, \sin\th \, Y^\ell Y_{,\th}^{*\ell'} = -(\ell + 1) f_{\ell -1} \mathcal{Q}_{\ell }+\ell  f_{\ell +1} \mathcal{Q}_{\ell +1} , \\
    \widebar{\Sc}_{2} f_{\ell} & \equiv f_{\ell'} \int \dd\Om \, \sin\th \cos\th \, Y^\ell Y_{,\th}^{*\ell'} = -(\ell +1) f_{\ell -2} \mathcal{Q}_{\ell -1} \mathcal{Q}_{\ell } +f_{\ell } \left(\ell  \mathcal{Q}_{\ell +1}^2-(\ell +1) \mathcal{Q}_{\ell }^2\right)+\ell 
   f_{\ell +2} \mathcal{Q}_{\ell +1} \mathcal{Q}_{\ell +2} , \\
    \Sc\widebar{\Sc} f_{\ell} & \equiv f_{\ell'} \int \dd\Om \, \sst \, Y_{,\th}^\ell Y_{,\th}^{*\ell'} = -(\ell -2) (\ell +1) f_{\ell -2} \mathcal{Q}_{\ell -1} \mathcal{Q}_{\ell
   } + f_{\ell } \left(\ell ^2 \mathcal{Q}_{\ell +1}^2+(\ell +1)^2 \mathcal{Q}_{\ell }^2\right) \notag \\
   & \hspace{120pt} -\ell  (\ell +3) f_{\ell +2} \mathcal{Q}_{\ell +1} \mathcal{Q}_{\ell +2} .
\end{align}
With the knowledge of these integrals we can compute the following operators, where we defined $\Sc_{0} f_{\ell} \equiv 0$ and $\C_0 f_{\ell} \equiv f_{\ell}$. For the class of equations belonging to the second group, one would need to use
\begin{align}
    \A_n f_\ell & \equiv f_{\ell'} \int \dd\Om \, \cos^n\!\th \left( Y_{,\th}^\ell Y_{,\th}^{*\ell'} + \frac{Y_{,\cf}^\ell Y_{,\cf}^{*\ell'}}{\sst} \right) = \left[ \C_n (\la+2) + n \, \Sc_n \right]f_{\ell} , \\
    \B_n f_\ell & \equiv f_{\ell'} \int \frac{\dd\Om}{\sin\th} \, \cos^n\!\th \left( Y_{,\cf}^\ell Y_{,\th}^{*\ell'} - Y_{,\th}^\ell Y_{,\cf}^{*\ell'} \right) = \ii m \, n  \,\C_{n-1} f_{\ell} , \\
    \X_n f_{\ell} & \equiv f_{\ell'} \int \dd\Om \, \cos^n\!\th \left( X^\ell Y_{,\th}^{*\ell'} - W^\ell Y_{,\cf}^{*\ell'} \right) = \ii m \left[ \C_n \left( \la - 2n \right) + 2 n \Sc_{n}\right] f_\ell , \\
    \widebar{\X}_n f_{\ell} & \equiv f_{\ell'} \int \dd\Om \, \cos^n\!\th \left( \sin\th \, W^\ell Y_{,\th}^{*\ell'} +  \frac{X^\ell Y_{,\cf}^{*\ell'}}{\sin\th} \right) = 2\left[ n m^2 \C_{n-1} -  (n+1) \Sc_{n+1} - \left(\frac{\widebar{\Sc}_{n+1}}{2} + \C_{n+1} \right) \frac{\ka_0}{\la} \right] f_{\ell} , \\
    \widetilde{\A}_2 f_{\ell} & \equiv f_{\ell'} \int \dd\Om \left( \sst \, Y_{,\th}^\ell Y_{,\th}^{*\ell'} - Y_{,\cf}^\ell Y_{,\cf}^{*\ell'} \right) = \left( \Sc\widebar{\Sc} - m^2 \right) f_{\ell} , \\
    \widetilde{\B}_2 f_{\ell} & \equiv f_{\ell'} \int \dd\Om \, \sin\th \left( Y_{,\cf}^\ell Y_{,\th}^{*\ell'} + Y_{,\th}^\ell Y_{,\cf}^{*\ell'} \right) = \ii m \left( \widebar{\Sc}_1 - \Sc_1  \right) f_{\ell} ,
\end{align}
while for those of the third group one has
\begin{align}
    \F_n f_{\ell} & \equiv f_{\ell'} \int \dd\Om \, \cos^n\!\th \left( Y_{,\th}^\ell X^{*\ell'} - Y_{,\cf}^\ell W^{*\ell'} \right) = \ii m \left[ \left( \la + 2n +4 \right)\C_n - 2\C_n (\la+2) -2n \Sc_n \right] f_\ell , \\
    \G_n f_{\ell} & \equiv f_{\ell'} \int \dd\Om \, \cos^n\!\th \left( \sin\th \, Y_{,\th}^\ell W^{*\ell'} + \frac{Y_{,\cf}^\ell X^{*\ell'}}{\sin\th} \right) = \left[ 2 m^2 n \C_{n-1} -\left( \la + 2n +4 \right) \Sc_{n+1} -2 \C_{n+1} (\la+2) \right] f_{\ell} , \\
    \Hs f_{\ell} & \equiv f_{\ell'} \int \dd\Om \, \sst \,  Y^\ell W^{*\ell'}  = \left[ (\la+2) \left( \C_2 - 1 \right) - 2 \widebar{\Sc}_2 + 2 m^2 \right] f_{\ell} , \\
    \widebar{\Hs} f_{\ell} & \equiv f_{\ell'} \int \dd\Om \, \sst \,  Y^\ell X^{*\ell'}  = -2\ii m \left( \widebar{\Sc}_1 - \C_1 \right) f_{\ell} , \\
    \J_0 f_{\ell} & \equiv f_{\ell'} \int \dd\Om \left( W_\ell W^{*\ell'} + \frac{X^\ell X^{*\ell'}}{\sst} \right)  = \ka_0  f_{\ell} , \\
    \J_2 f_{\ell} & \equiv f_{\ell'} \int \dd\Om \, \cct \left( W_\ell W^{*\ell'} + \frac{X^\ell X^{*\ell'}}{\sst} \right)  = \left[ 8m^2  + \left((\la - 2) \C_2 - 2\widebar{\Sc}_2 -2 \right) (\la+2) - 2 (\la + 6) \Sc_2  \right] f_{\ell} , \\
    \K_0 f_{\ell} & \equiv f_{\ell'} \int \frac{\dd\Om}{\sin\th} \left( W_\ell X^{*\ell'} - X^\ell W^{*\ell'} \right)  = 0 , \\
    \K_2 f_{\ell} & \equiv f_{\ell'} \int \frac{\dd\Om}{\sin\th} \cct \left( W_\ell X^{*\ell'} - X^\ell W^{*\ell'} \right)  = -2\ii m \left[ \left(\widebar{\Sc}_1 +3 \C_1 \right) (\la+2) + (\la+6) \left( \Sc_1 - \C_1 \right) \right] f_{\ell} ,
\end{align}
\end{widetext}
where we remind that $\la = \ell^2 + \ell - 2$ and $\ka_0 = \la(\la+2)$.

\section{Decoupling coefficients}\label{app:coefficients}
Let us present the explicit form of the coefficients that appear in equations~\eqref{eq:RW_redef_h1_2nd_order}--\eqref{eq:Zer_redef_H1_2nd_order} and that allow a complete decoupling of the slow-spinning Regge-Wheeler and Zerilli equations. We report here only the coefficients proportional to $a$, while those proportional to $a^2$ are reported in the supplemental material
\begin{widetext}
\begin{align}
    c_0^\ell & = -\frac{6f_0}{\ell(\ell+1) r^3}, \qquad c_1^\ell = \frac{1}{r^2 f_0}, \qquad c_K^\ell = \frac{f_0}{\rho r} c_H^\ell + \frac{\rho^2}{r^2(\la+2)} - \frac{24 +3r(\la-6) + 8r^2\la + r^3 \la^2(\la+6)}{4r^6(\la+2)(3+\la r)} \\
    c_H^\ell & = \frac{21+18r\la + r^2 \la(3\la-2)}{r f_0 (\la+2)(3+r \la)^2} \rho^3 + \frac{12r-11}{4r^4f_0}\rho \notag \\
    & + \frac{8 \la ^3 r^5-2 \la ^2 (19 \la +20) r^4+\la  \left(29 \la ^2-76 \la -312\right) r^3+12 \left(9 \la ^2+22 \la -36\right)
   r^2+27 (\la +22) r-180}{2 (\la +2) r^5 f_0 (3 + \la  r)^3} \rho \\
   s_{0,1}^\ell & = - \frac{1}{\ell+1}, \qquad t_{0,1}^\ell = \frac{2\rho r}{f_0(\ell+1)^2} \qquad  s_{0,0}^\ell = \frac{f_0}{(\ell^2 r + 3\ell r+3)} \left[ \frac{2(\ell+3) r^2\rho^2}{(\ell+1)^2} - \frac{(\ell+2)(2 r \ell ^2+6 r \ell -\ell +3)}{2r(\ell+1)} \right] \\
   t_{0,0}^\ell & = \frac{\rho}{(\ell^2 r + 3\ell r+3)} \left[ \frac{-4 r^2 \ell  \left(\ell ^2+5 \ell +6\right)+2 r \left(\ell ^3+5 \ell ^2+2 \ell -6\right)+3 (\ell +1)}{2r(\ell+1)^2} - \frac{2(\ell+3) r^3 \rho^2}{(\ell+1)^2}\right] \\
   s_{0,K}^\ell & = \frac{2(\ell+3) f_0}{(\ell+1) \rho r} \left[ \frac{\ell(\ell r+2r-1)}{r^3} + \frac{2\rho^2}{\ell+1} \right] \qquad t_{0,K}^\ell = \frac{2(\ell+3)(\ell+2)}{(\ell+1)r^2} \\
   s_{0,H}^\ell & = \frac{\ell(\ell+3)\left[4(\ell+2)r-3(\ell+3) \right]-4}{(\ell+1)^2r^3} + \frac{4(\ell+3)\rho^2}{(\ell+1)^2} \qquad t_{0,H}^\ell = \frac{2(\ell+3)(2\ell r+4r-3)}{(\ell+1)^2 f_0 r^2} \rho
\end{align}

\end{widetext}

\section{Effective potential}\label{app:potentials}

In section~\ref{sec:metric&perturbation} we found that the perturbations of a slowly-rotating Kerr BH can be recast into the same form of the Regge-Wheeler and Zerilli equations and that the potentials receive a correction in the spin of the form given in equation~\eqref{eq:potential_slow_rot}. Here, we report the explicit form of $\widebar{V}^\ell_{(\pm),2}$ defined in equation~\eqref{eq:potential_2}. For the axial sector we have
\begin{widetext}
\begin{align}
    \widebar{V}_{(-),2}^\ell = & \, - \frac{m^2}{r^4} + m^2 f_0 \Bigg[ - \frac{24(7-6r)}{(\la + 2)^3r^6}  - \frac{12(47-40r)}{(\la+2)^2r^6} + \frac{2(6r^2-250r-315)}{(\la+2)r^6} + \frac{420(6-5r)}{(5+4\la)r^6} \Bigg]  \notag \\
    & \, - m^2 f_0 \left[ \frac{2(\la-10)}{(\la+2)(5+4\la)r^2} \rho^2 + 3\frac{6r^2(4\la-19) - 26r(\la-13) - 231}{(\la+2)^2r^{10}\rho^2}  \right] ,
\end{align}
while for the polar sector we find
\begin{align}
   \widebar{V}_{(+),2}^\ell = & \, -\frac{m^2}{r^4} + \frac{m^2 f_0}{(3 + r \la)^5} \Bigg[ \frac{2 (4 \la -1) \la ^4 r}{\la +2} -\frac{2 \left(34 \la ^5+135 \la ^4-513 \la ^3-1753 \la ^2-312 \la +60\right) \la ^3}{(\la +2)^3 (5+4 \la)} \notag \\
   & \, -\frac{\left(70
   \la ^5+810 \la ^4+5599 \la ^3-398 \la ^2-26676 \la +1704\right) \la ^3}{(\la +2)^3 (5+4 \la) r} \notag \\
   & \, -\frac{\left(929 \la ^5+5310
   \la ^4+40121 \la ^3+33516 \la ^2-128980 \la +16272\right) \la ^2}{(\la +2)^3 (5+4 \la) r^2} \notag \\
   & \, -\frac{3 \left(2993 \la ^5+8760 \la
   ^4+93012 \la ^3+146824 \la ^2-243912 \la +19728\right) \la }{2 (\la +2)^3 (5+4 \la) r^3}-\frac{1944 \left(8 \la ^2-27 \la
   -134\right)}{(\la +2)^3 (5+4 \la) r^7} \notag \\
   & \, -\frac{9 \left(6403 \la ^3-20642 \la ^2-54548 \la +135720\right)}{2 (\la +2)^3 (5+4 \la) r^6}-\frac{9
   \left(5375 \la ^4-11512 \la ^3+22328 \la ^2+229320 \la -65664\right)}{2 (\la +2)^3 (5+4 \la) r^5} \notag \\
   & \, -\frac{3 \left(8503 \la ^5+1374 \la
   ^4+160020 \la ^3+455216 \la ^2-349416 \la +11664\right)}{2 (\la +2)^3 (5+4 \la) r^4} \Bigg] \notag \\
   & \, + \frac{2 m^2 f_0 \rho^2}{(\la+2)(5+4\la)(3+ r \la)^4} \Bigg[ \frac{3 \left(3 \la ^4+26 \la ^3+214 \la ^2+528 \la -24\right) \la }{(\la +2)^2}- \left(\la ^2-6 \la +2\right) \la ^3 r^2 \notag \\
   & \, +\frac{18
   \left(17 \la ^2-64 \la -304\right)}{(\la +2)^2 r^3}+\frac{6 \left(71 \la ^3-136 \la ^2-754 \la +468\right)}{(\la +2)^2 r^2} +\frac{12 \left(19 \la ^3+16 \la ^2+4 \la +312\right) \la
   }{(\la +2)^2 r} \notag \\
   & \, -\frac{4 \left(2
   \la ^4-7 \la ^3-57 \la ^2-76 \la +12\right) \la ^2 r}{(\la +2)^2} \Bigg] \notag \\
   & \, + \frac{m^2 f_0}{(\la+2)^2(3+ r \la)^4 \rho^2} \Bigg[ -\frac{3 \left(13 \la ^2+61 \la +36\right) \la ^5}{r^2}+\frac{\left(92 \la ^3+197 \la ^2-942 \la -1032\right) \la ^4}{r^3} \notag \\
   & \, +\frac{\left(-107
   \la ^4+665 \la ^3+4582 \la ^2+726 \la -7272\right) \la ^3}{2 r^4}-\frac{3 \left(118 \la ^4+138 \la ^3-1133 \la ^2-3180 \la
   +1872\right) \la ^2}{r^5} \notag \\
   & \, -\frac{18 \left(84 \la ^3-137 \la ^2+317 \la +90\right) \la }{r^7}-\frac{3 \left(291 \la ^4+455 \la ^3+2211 \la
   ^2-5106 \la +1080\right) \la }{r^6} \notag \\
   & \, +\frac{54 \left(115 \la ^2-1138 \la +918\right)}{r^9}-\frac{9 \left(443 \la ^3+2319 \la ^2-8670 \la
   +2808\right)}{2 r^8}+\frac{648 (42 \la -97)}{r^{10}}+\frac{25920}{r^{11}} \Bigg] .
\end{align}
\end{widetext}

\section{The superpotential at second-order in the spin}\label{app:generators}
In section~\ref{sec:isospectrality}, we showed how to the derive the superpotential that generates the first-order correction to the Regge-Wheeler and Zerilli equations. Here, we proceed to show the analogous calculation at second-order in the spin. First of all, it is convenient to split $W_2^\ell$ and $\ka_2$ in five parts, such as
\begin{align}
    \label{eq:superpotential_2ord}
    W_2^\ell & = \widetilde{W}_{2}^\ell + \widebar{W}_{2}^\ell \rho^2 + m^2 \left( \frac{\widehat{W}_{2}^\ell}{\rho^2} + \widecheck{W}_{2}^\ell + \wideparen{W}_{2}^\ell \rho^2 \right), \\
    \ka_2 & = \widetilde{\ka}_{2} + \widebar{\ka}_{2} \rho^2 + m^2 \left( \frac{\widehat{\ka}_{2}}{\rho^2} + \widecheck{\ka}_{2} + \wideparen{\ka}_{2} \rho^2 \right) ,
\end{align}
as well as splitting the tortoise coordinate function at second-order as
\begin{equation}\label{eq:Tortoise2_isospectrality}
    f_2 = \widetilde{f}_2 + m^2 \left( \frac{\widehat{f}_2 + f_1^2}{\rho^2} + \widecheck{f}_2 \right) .
\end{equation}
Let us insert this expansion in equations~\eqref{eq:Darboux_deriv}--\eqref{eq:Darboux} and take quadratic terms in the spin, assuming that $W_0^\ell$, $W_1^\ell$, $f_1$ and $\ka_1$ are specified as in section~\ref{sec:isospectrality}. By setting to zero the term proportional to $\rho^2 m^0$ in equation~\eqref{eq:Darboux_deriv}, we obtain
\begin{align}
    \widebar{W}_2^\ell = & \, \widebar{k}_2 + \frac{(\la -6) (2 \la +1)}{3 (\la +2) (4 \la +5) r (\la  r+3)} + \frac{2}{3 r f_0(\la+3)} \notag \\
    & \, + \frac{9 \left(\la ^2+\la -6\right)+\left(-4 \la ^3+34 \la +36\right) r^2}{3 (\la +2) (\la +3) (4 \la +5) r^2 (\la  r+3)} .
\end{align}
Analogously to the case linear in $a$, we find that the $\rho^2 m^0$ term of equation~\eqref{eq:Darboux} vanishes for this choice of the tortoise coordinate.
\begin{align}\label{eq:tortoise_m0}
    \widetilde{f}_2 = & \, \frac{\ka_0 \widebar{\ka}_2}{2} + \frac{1}{r^2 f_0} + \frac{f_0(\la+5)}{r(4\la+5)} \notag \\
    & \, + \left( \frac{\ka_0}{2} +\be_0^2 W_0^\ell \right)\widebar{W}_2^\ell - \frac{3}{2} \frac{\dd \widebar{W}_2^\ell}{\dd r_{*,0}}.
\end{align}
We now move at the term proportional to $\rho^0 m^0$, and equations~\eqref{eq:Darboux_deriv}--\eqref{eq:Darboux} are satisfied simultaneously when
\begin{widetext}
\begin{align}
    & \widetilde{W}_2^\ell = \widetilde{k}_2 \left[ 1 + \frac{\be_0^2 W_0^\ell}{\ka_0 } \right] + \frac{1}{6 r^3 f_0 (\la + 3)} + \frac{1}{(5+4\la)(3+r\la)^3} \bigg[ \frac{\left(8 \la ^3+52 \la ^2+95 \la +54\right) \la ^2}{6 (\la +3)} -\frac{54 (\la
   -2)}{(\la +2) r^6} \notag \\
    & -\frac{9 \left(31 \la ^2-111 \la +326\right)}{4 (\la +2) r^5} -\frac{3
   \left(47 \la ^3-368 \la ^2+843 \la -1074\right)}{4 (\la +2) r^4} +\frac{-100 \la ^4+1567 \la ^3-3045 \la ^2+7836 \la -1116}{12 (\la
   +2) r^3}  \notag \\
   & - \frac{4 \la ^6-148 \la ^5-159 \la ^4-393 \la ^3-4095 \la ^2-252 \la -324}{6 (\la +2) (\la +3)r^2} +\frac{\left(12 \la ^5+34 \la ^4+281 \la ^3+1029 \la ^2+618 \la +216\right) \la }{6 (\la +2) (\la +3) r} \bigg] , \\
   & \widebar{k}_2 = \widebar{k}_1\frac{\la(\la-10)}{5+4\la}, \qquad \widebar{\ka}_2 = - \widebar{k}_2 + \widetilde{k}_2 \left(\frac{6}{\ka_0}\right)^2, \qquad \widetilde{\ka}_2 = - \widetilde{k}_2 .
\end{align}
We can see from the previous equation that the constant $\widetilde{k}_2$ that comes from the integration of $\widetilde{W}_2^\ell$ is not fully specified by the equation. We now move to solve the terms proportional to $m^2$ in the Darboux relations. Let us start by checking the relation~\eqref{eq:Darboux_deriv} for the term proportional to $\rho^2m^2$, for which we infer
\begin{align}
    \wideparen{W}_2^\ell = & \, \wideparen{k}_2 + \frac{1}{(3+r \la)^3} \Bigg[ \frac{-23 \la ^4+116 \la ^3+590 \la ^2+352 \la +72}{3 (\la +2)^3 (4 \la +5)}+\frac{2 (2 \la +1) \la ^2 r^2}{3 (\la +2) (4 \la
   +5)}+\frac{-17 \la ^2+64 \la +304}{(\la +2)^3 (4 \la +5) r^2} \notag \\
   & \, +\frac{2 \left(-\la ^4+17 \la ^3+78 \la ^2+80 \la +24\right) \la  r}{3
   (\la +2)^3 (4 \la +5)}+\frac{-28 \la ^3+38 \la ^2+320 \la +48}{(\la +2)^3 (4 \la +5) r} \Bigg] ,
\end{align}
and again $\wideparen{k}_2$ is an integration constant. From requiring consistency with equation~\eqref{eq:Darboux} we must fix
\begin{align}
    \widecheck{f}_2 = & \, \frac{\ka_0 \wideparen{\ka}_2}{2} + \frac{f_0(\la-10)}{r^2(\la+2)(5+4\la)} - \frac{\be_0^2}{2}\widebar{W}_1^\ell + \left( \frac{\ka_0}{2} +\be_0^2 W_0^\ell \right)\wideparen{W}_2^\ell - \frac{3}{2} \frac{\dd \wideparen{W}_2^\ell}{\dd r_{*,0}}.
\end{align}
Then repeating the procedure for the terms proportional to $m^2\rho^0$, we obtain
\begin{align}
    & \widecheck{W}_2^\ell = \widecheck{k}_2 + \frac{(\wideparen{k}_2 + \wideparen{\ka}_2)\ka_0 W_0^\ell}{2r} + \be_0^2 \frac{ \wideparen{\ka}_2 {W_0^\ell}^2}{2r^2} + \frac{1}{ (3+r \la)^4} \Bigg[ -\frac{2}{9} \la ^5 r^2+\frac{3 \left(80 \la ^2-217 \la -1186\right)}{2 (\la +2)^3 (4 \la +5) r^6} \notag \\
    & \, + \frac{\left(284 \la ^6+1051 \la ^5+332 \la ^4-5822 \la ^3-11860 \la ^2-2120 \la +720\right) \la ^2}{36 (\la +2)^3 (4 \la
   +5)} +\frac{811 \la ^3-6416 \la ^2-9686
   \la +32652}{4 (\la +2)^3 (4 \la +5) r^5} \notag \\
   & \, +\frac{397 \la ^4-3527 \la ^3-782 \la ^2+21828 \la -14568}{2 (\la +2)^3 (4 \la +5)
   r^4}+\frac{527 \la ^5-3322 \la ^4+3296 \la ^3+22592 \la ^2-36296 \la +4176}{4 (\la +2)^3 (4 \la +5) r^3} \notag \\
   & \, +\frac{776 \la ^6-1123 \la
   ^5+10464 \la ^4+25902 \la ^3-39724 \la ^2+18600 \la +432}{12 (\la +2)^3 (4 \la +5) r^2} +\frac{\left(8 \la ^4-46 \la ^3-202 \la ^2-156
   \la +9\right) \la ^3 r}{9 (\la +2) (4 \la +5)} \notag \\
   & \, +\frac{\left(164 \la ^6+518 \la ^5+2379 \la ^4+3384 \la ^3-2974 \la ^2+3432 \la
   +216\right) \la }{6 (\la +2)^3 (4 \la +5) r} \Bigg] ,
\end{align}
as well as
\begin{align}
    \widehat{f}_2 = & \, \frac{1}{4}\frac{\dd ^2 \widecheck{f}_2}{\dd r_{*,0}^2} + \left[W_0^\ell \left( \ka_0 + \be_0^2 W_0^\ell \right) + \frac{3}{2}\frac{\dd W_0^\ell}{\dd r_{*,0}} \right] \widecheck{f}_2 + \frac{\ka_0 \widecheck{\ka}_2}{2} + f_1 \left[ \frac{2}{r^3} + \frac{3}{2}\frac{\dd \widebar{W}_1^\ell}{\dd r_{*,0}} \right]  + \left( \frac{\ka_0}{2} +\be_0^2 W_0^\ell \right)\widecheck{W}_2^\ell - \frac{3}{2} \frac{\dd \widecheck{W}_2^\ell}{\dd r_{*,0}} \notag \\
    & \, - \widebar{W}_1^\ell \widetilde{W}_1^\ell + \frac{r(\la-10) + 12}{2r^5(\la+2)} + f_0 \frac{3 \left(61 \la ^2+74 \la -180\right)+2 \left(25 \la ^3+45 \la ^2+96 \la +320\right) r}{(\la +2)^3 (5 + 4 \la) r^6} .
\end{align}
Finally from the terms proportional to $m^2 \rho^{-2}$ we obtain
\begin{align}
    & \widetilde{W}_2^\ell = \frac{(\widecheck{k}_2 + \widecheck{\ka}_2)\ka_0}{36}\left( \ka_0 + 2 \be_0^2 W_0^\ell \right) + \frac{f_0}{(3+r \la)^5} \Bigg[\frac{(\la +10)^2 \la ^5}{36 (\la +2) r} +  \frac{\left(\la ^3+60 \la ^2+302 \la +624\right) \la ^4}{12 (\la +2)^2 r^2} \notag \\
   & \, +\frac{\left(7 \la ^3+322 \la ^2+158 \la +1202\right) \la
   ^3}{6 (\la +2)^2 r^3} +\frac{\left(21 \la ^3+1618 \la ^2-1184 \la +1532\right) \la ^2}{4 (\la +2)^2 r^4}+\frac{3 \left(9 \la ^2+5777 \la
   -3426\right) \la }{4 (\la +2)^2 r^6} \notag \\
   & \, +\frac{3 \left(8 \la ^3+2299 \la ^2-1910 \la +400\right) \la }{4 (\la +2)^2 r^5} -\frac{27 (8 \la
   -259)}{2 (\la +2)^2 r^8}+\frac{9 (1337 \la -360)}{2 (\la +2)^2 r^7}-\frac{162}{(\la +2)^2 r^9} \Bigg], \\
   & \widetilde{\ka}_2 = (\widecheck{k}_2 + \widecheck{\ka}_2) \left( \frac{\ka_0}{6} \right)^2, \qquad \wideparen{k}_2 = \widebar{k}_1 \frac{60 + 34\la + 5\la^2}{(\la+2)^2(5+4\la)}, \qquad \wideparen{\ka}_2 = \left(\frac{6}{\ka_0}\right)^2 \widecheck{k}_2 - \widebar{k}_1 \frac{80 + 60 \la + 13 \la^2}{(\la+2)^2(5+4\la)} .
\end{align}
With this, we completely specify the function $W_2^\ell$, upon choice of the integrating constants. Indeed, we see again how the constants $\widecheck{k}_2 $ and $ \widecheck{\ka}_2$ are not specified by the validity of the Darboux transformation.
\end{widetext}

\section{Transformation theory at second-order in the spin}\label{app:transformation}
We now sketch the form of the calculations carried to obtain the second-order spin correction functions $\f_2$, $T_2$, $\be_2$ and $\widetilde{R}_2 \equiv R_2 - Q_2$ and constant $K_2$, which appear in the transformations defined in equations~\eqref{eq:transformation_U}, \eqref{eq:transformation_LaU} and~\eqref{eq:rel5}. We found convenient to perform the following decomposition in powers of $\rho$ and $m$
\begin{align}
    F_2 & = \sum_{i=0}^2 F_{2,i} \rho^i + m^2 \sum_{i=-2}^2 \widebar{F}_{2,i} \rho^i ,
\end{align}
where $F_2$ stands for any of the functions $\f_2$, $T_2$, $\be_2$ and $\widetilde{R}_2$ (as well $R_2$) and constant $K_2$.
Moreover, in order to express the following formulae in a compact form, it is useful to spilt the Regge-Wheeler/Zerilli potentials, the Teukolsky potential and reduced separation constant as\footnote{Note that the slightly different definition of $Q_2$ compared with that given in equation~\eqref{eq:VT2_format} comes from making explicit the $m^2$ dependency in $\la_2^{(s)}$.}
\begin{align}
    & V_2 = V_{2,0} + V_{2,2} \rho^2 + m^2 \left( \frac{\widebar{V}_{2,-2}}{\rho^2} + \widebar{V}_{2,0} + \widebar{V}_{2,2} \right) , \\
    & Q_2 = \sum_{i=0}^2 Q_{2,i} \rho^i + m^2 \sum_{i=-2}^2 \widebar{Q}_{2,i} \rho^i , \\
    & \la_2^{(s)} = \widetilde{\la}_2 + m^2 \widebar{\la}_2 .
\end{align}
Finally, we make extensive use of the super-potential $W_2$ and the tortoise coordinate $f_2$ as defined in equations~\eqref{eq:superpotential_2ord} and~\eqref{eq:Tortoise2_isospectrality} respectively.
We proceed to solve the relations $\Rel_1 = 0$, $\Rel_2 = 0$, $\Rel_4 = 0$ and $\Rel_5 = 0$ at each order in $\rho$ and $m$. From the term proportional to $\rho^3 m^0$ we find
\begin{equation}
    \f_{2,2} = \be_{2,2} = \widetilde{R}_{2,2} = 0.
\end{equation}
We can manipulate the terms proportional to $\rho^2m^0$ into three algebraic relations and one first-order differential equation, whose solutions are
\begin{align}
    & 2\f_{2,1} = T_{2,2} + \widetilde{T}_0\frac{V_{2,2}}{V_0} + \frac{1}{V_0}\frac{\dd V_{2,2}}{\dd r_{*,0}} , \\
    & 2\be_{2,1} = K_{2,2} - \frac{1}{W_0}\left( \frac{Q_{2,2}V_0}{Q_0} + V_{2,2} \right) - \be_0 T_{2,2} , \\
    & 2\widetilde{R}_{2,1} = Q_0 T_{2,2} + \left(\!\widetilde{T}_0+2\frac{3-2r}{r^2}\right)Q_{2,2} - \frac{\dd Q_{2,2}}{\dd r_{*,0}} , \\
    & T_{2,2} = \frac{2\la-5}{r(5+4\la)} - \frac{\widetilde{\la}_2}{r} - (3+\be_0)\widebar{W}_2 + t_{2,2} ,
\end{align}
where $t_{2,2}$ is an integration constant.
The structure of the equations is maintained at order $\rho^1 m^0$, for which we get
\begin{align}
    2\f_{2,0} = & \, T_{2,1} + \frac{Q_0 W_0}{V_0}\be_{2,1} + \frac{1}{V_0} \,\frac{\dd V_0 \f_{2,1}}{\dd r_{*,0}} , \\
    2\be_{2,0} = & \, K_{2,1} -\be_0 T_{2,1} - \widetilde{T}_0 \be_{2,1} \notag \\
    & \, - \frac{V_0}{W_0} \left(\frac{R_{2,1}}{Q_0} + \f_{2,1} \right), \\
    2\widetilde{R}_{2,0} = & \, Q_0 T_{2,1} + \widetilde{T}_0 Q_{2,1} + 2 \frac{3-2r}{r^2} R_{2,1} \notag \\
    & \, + Q_0 W_0 \be_{2,1} - \frac{\dd R_{2,1}}{\dd r_{*,0}} , \\
    2T_{2,1} = & \, \frac{2}{r^3}\left(2r - \frac{1}{f_0 r}\right) - \ka_0 \widebar{\ka}_2 + \frac{2f_0 \widetilde{\la}_2}{r^2} \notag \\
    & \, + T_{2,2} \widetilde{T}_0 - V_{2,2} - Q_{2,2} + 2 \widetilde{f}_2 + t_{2,1},
\end{align}
where $t_{2,1}$ is another integration constant.
We find that the equations at order $\rho^0 m^0$ are consistent only by selecting $\widetilde{f}_2$ as given in equation~\eqref{eq:tortoise_m0}. Explicitly we get 
\begin{align}
    T_{2,0} = & \, \frac{2r-1}{f_0 r^4} - \frac{\be_0 W_0}{f_0 r^2} - 2 \widetilde{f}_2 \left( 4 \frac{3-2r}{r^2} +\widetilde{T}_0 \right) \notag \\
    & \, - \frac{1}{Q_0} \left[ 2 \frac{3-2r}{r^2} R_{2,0} + Q_{2,0} \widetilde{T}_0 \right] - \be_{2,0} W_0 \notag \\
    & \, + \frac{1}{Q_0^2} \frac{\dd \widetilde{f}_2 \, Q_0^2}{\dd r_{*,0}} - \frac{1}{Q_0} \frac{\dd R_{2,0}}{\dd r_{*,0}} ,\\
    K_{2,2} = & \, 2\be_0 t_{2,2} - \frac{4\la(\la-10)}{5+4\la}, \\
    K_{2,1} = & \, 2\be_0 t_{2,1} + \ka_0 t_{2,2} + \frac{36 \be_0 \widetilde{k}_2}{\ka_0} \\
    K_{2,0} = & \, \ka_0 t_{2,1} + 2 \be_0^2 \widetilde{k}_2 .
\end{align}
Also in the first-order case, the integration constants $t_{2,1}$ and $t_{2,2}$ are not specified by the transformation.
Let's now move to the terms proportional to $m^2$. The procedures follows the same lines of the previous one. From the terms proportional to $\rho^3 m^2$ we obtain
\begin{equation}
    \widebar{\f}_{2,2} = \widebar{\be}_{2,2} = \widebar{\widetilde{R}}_{2,2} = 0.
\end{equation}
Then, from the terms proportional to $\rho^2 m^2$ we get
\begin{align}
    2\widebar{\f}_{2,1} = & \, \widebar{T}_{2,2} + \frac{\dd}{\dd r_{*,0}} \left( \frac{\widebar{V}_{2,2}}{V_0} + \frac{1}{2} \frac{V_{1,2}^2}{V_0^2} \right)  \notag \\
    & \, - \frac{Q_0 W_0}{V_0^2} \left( \widebar{V}_{2,2} V_0 + V_{1,2}^2 \right), \\
    2\widebar{\be}_{2,1} = & \widebar{K}_{2,2} -  \frac{\widebar{Q}_2 V_0 - Q_{2,1} V_{2,1}}{Q_0 W_0} \! + \! \frac{\widebar{V}_{2,2}}{W_0}  -\be_0 \widebar{T}_{2,2} , \\
    2\widebar{\widetilde{R}}_{2,1} = & \, Q_0 \widebar{T}_{2,2} + \left( \widetilde{T}_0 + 2 \frac{3-2r}{r^2} \right)\widebar{Q}_{2,2} \notag \\
    & \,  - T_{1,1} Q_{1,1} - \frac{\dd Q_{2,2}}{\dd r_{*,0}} , \\
    \widebar{T}_{2,2} = & \, \widebar{t}_{2,2} - \frac{2(\la-10)}{r(\la+2)(5+4\la)} - \frac{\widebar{\la}_2}{r} \notag \\
    & \, - (3+\be_0)\wideparen{W}_2 , 
\end{align}
From the terms proportional to $\rho^1 m^2$ we get
\begin{align}
    2\widebar{\f}_{2,0} = & \, \widebar{T}_{2,1} + \frac{Q_0 W_0}{V_0} \left( \widebar{\be}_{2,1} + \frac{V_{1,2}}{V_0}\be_{1,1} \right) \notag \\
    & \, + \frac{1}{V_0} \frac{\dd V_0 \, \widebar{\f}_{2,1}}{\dd r_{*,0}}- \f_{1,1} \frac{\dd}{\dd r_{*,0}} \frac{V_{1,2}}{V_0} , \\
    2\widebar{\be}_{2,0} = & \, \widebar{K}_{2,1} - \be_0 \widebar{T}_{2,1} - \widetilde{T}_0 \widebar{\be}_{2,1} + T_{1,2} \be_{1,1} \notag \\
    & \,  - \frac{V_0}{W_0} \left[ \f_{1,1} \left( \frac{V_{1,2}}{V_0} + \frac{Q_{2,1}}{Q_0} \right) - \widebar{\f}_{2,1} \right]  \notag \\
    & \, + \frac{R_{1,1} V_{1,2} - \widebar{R}_{2,1} V_0}{W_0 Q_0}  , \\
    2\widebar{\widetilde{R}}_{2,0} = & \, Q_0 \widebar{T}_{2,1} + \widebar{Q}_{2,1} \widetilde{T}_0 - Q_{1,2} T_{1,1} - Q_{1,1} T_{1,2}  \notag \\
    & \, + 2 \frac{3-2r}{r^2} \widebar{R}_{2,1} + Q_0 W_0 \widebar{\be}_{2,1} - \frac{\dd \widebar{R}_{2,1}}{\dd r_{*,0}} , \\
    2 \widebar{T}_{2,1} = & \, \widebar{t}_{2,1} -\ka_0\wideparen{\ka}_2 + \frac{2f_0 \widebar{\la}_2}{r^2} + \widebar{T}_{2,2}\widetilde{T}_0 - \frac{T_{1,2}^2}{2}  \notag \\
    & \, - \widebar{V}_{2,2} - \widebar{Q}_{2,2} + 2 \wideparen{f}_2 , 
\end{align}
From the terms proportional to $\rho^0 m^2$ we get
\begin{widetext}
\begin{align}
    2\widebar{\f}_{2,-1} = & \, \widebar{T}_{2,0} + \left(T_{1,2} - 2\f_{1,1}\right) f_1 + \wideparen{f}_2 \widetilde{T}_0 + \frac{Q_0 W_0}{V_0} \left[ \widebar{\be}_{2,0} + \frac{V_{1,2}}{V_0}\be_{1,1} - \be_0 \left( \frac{V_{1,2}V_{1,0}}{V_0^2} + \frac{\widebar{V}_{2,0}}{V_0} \right) \right] \notag \\
    & \, -\f_{1,0} \frac{\dd}{\dd r_{*,0}}\frac{V_{1,2}}{V_0} + \frac{\dd}{\dd r_{*,0}}\frac{V_{1,2}V_{1,0}}{V_0^2} + \frac{\dd}{\dd r_{*,0}}\frac{V_{2,0}}{V_0} + \frac{1}{V_0} \frac{\dd V_0 \widebar{\f}_{2,0}}{\dd r_{*,0}}, \\
    2\widebar{\be}_{2,-1} = & \, \widebar{K}_{2,0} - \be_0 \widebar{T}_{2,0} - \widetilde{T}_0 \widebar{\be}_{2,0} + T_{1,2}\be_{1,0} + T_{1,1} \be_{1,1} + \frac{Q_{1,2} V_{1,0} + R_{1,0} V_{1,2} - Q_0 \widebar{V}_{2,0} - \widebar{R}_{2,0} V_0 }{W_0 Q_0} \notag \\
    & \, + \frac{V_0}{W_0} \left[ \f_{1,0} \left( \frac{V_{1,2}}{V_0} + \frac{Q_{1,2}}{Q_0} \right) + \frac{R_{1,1}}{Q_0}\f_{1,1} - \widebar{\f}_{2,0} + f_1 \left( \frac{V_{1,2}}{V_0} - \frac{Q_{1,2}}{Q_0} \right) -\wideparen{f}_2 \right] , \\
    2\widebar{\widetilde{R}}_{2,-1} = & \, Q_0 \widebar{T}_{2,0} - Q_{1,0} T_{1,2} - Q_{1,2} T_{1,0} - Q_{1,1} T_{1,1} + \widebar{Q}_{2,0} \widetilde{T}_0  + \left( Q_{1,2} \widetilde{T}_0 - Q_0 T_{1,2} - 2 \widetilde{R}_{1,1} \right) f_1 + Q_0 \wideparen{f}_2 \notag \\
    & \, + 2 \frac{3-2r}{r^2}\widebar{R}_{2,0} + Q_0 W_0 \widebar{\be}_{2,0} - Q_0 W_0^2 \frac{\dd}{\dd r_{*,0}} \frac{\wideparen{f}_2}{W_0^2} - Q_{1,2} \frac{\dd f_1}{\dd r_{*,0}} - \frac{\dd \widebar{R}_{2,0}}{\dd r_{*,0}} , 
\end{align}
whereas the expression for $\widebar{T}_{2,0}$ is extremely long and uninformative and we do not display it. Then, by setting to zero the terms proportional to $\rho^{-1}m^2$ we obtain
\begin{align}
    2\widebar{\f}_{2,-2} = & \, \widebar{T}_{2,-1} + \left( T_{1,1} - 2 \f_{1,0} \right) f_1 + \frac{Q_0 W_0}{V_0} \left( \widebar{\be}_{2,-1} + \frac{V_{1,0}}{V_0} \be_{1,1} \right) - \f_{1,1} \frac{\dd}{\dd r_{*,0}} \frac{V_{1,0}}{V_0} + \frac{1}{V_0} \frac{\dd V_0 \widebar{\f}_{2,-1}}{\dd r_{*,0}} , \\
    2\widebar{\be}_{2,-2} = & \, \widebar{K}_{2,-1} - \be_0 \widebar{T}_{2,-1} - \widetilde{T}_0 \widebar{\be}_{2,-1} + T_{1,1} \be_{1,0} + T_{1,0} \be_{1,1} + \frac{V_0}{W_0} \left[ \f_{1,1} \left( \frac{R_{1,0}}{Q_0} + \frac{V_{1,0}}{V_0} + f_1 \right) - \widebar{\f}_{2,-1} \right] \notag \\
    & \, + \frac{V_0}{W_0 Q_0} \left[ R_{1,1}\left( \frac{V_{1,0}}{V_0} + \f_{1,0} - f_1 \right) - \widebar{R}_{2,-1} \right], \\
    2\widebar{\widetilde{R}}_{2,-2} = & \, Q_0 \widebar{T}_{2,-1} - Q_{1,0} T_{1,1} - Q{1,1} T_{1,0} + \widebar{Q}_{2,-1} \widetilde{T}_0 + \left( Q_{1,1} \widetilde{T}_0 - Q_0 T_{1,1} - 2 \widetilde{R}_{1,0} \right) f_1 + 2 \frac{3-2r}{r^2}\widebar{R}_{2,-1} \notag \\
    & \, + Q_0 W_0 \widebar{\be}_{2,-1} - R_{1,1} \frac{\dd f_1}{\dd r_{*,0}} - \frac{\dd \widebar{R}_{2,-1}}{\dd r_{*,0}}, 
\end{align}
as well as $\widebar{T}_{2,-1}$ whose expression is extremely long and uninformative and we do not display it. Finally, from the terms proportional to $\rho^{-2}m^2$ we find
\begin{align}
    \widebar{K}_{2,2} = & \, 2\be_0\widebar{t}_{2,2} - \frac{4\la(60+34\la+5\la^2)}{(\la+2)^2(5+4\la)} , \\
    \widebar{K}_{2,1} = & \, 2\be_0\widebar{t}_{2,1} + \ka_0\widebar{t}_{2,2} + \frac{36\be_0 \widecheck{k}_2}{\ka_0} +4t_2\la - \frac{\be_0 t_2^2}{2} , \\
    \widebar{K}_{2,0} = & \, 2\be_0\widebar{t}_{2,0} + \ka_0 \left( \widebar{t}_{2,1} - \frac{t_2^2}{4} \right) + 18 \widecheck{k}_2 + 4t_1\la - \be_0 t_1 t_2 , \\
    \widebar{K}_{2,-1} = & \, 2\be_0 \widebar{t}_{2,-1} - \frac{\be_0 t_1^2}{2} + \ka_0 \left[ \widebar{t}_{2,0} +  \be_0 \left( \widecheck{k}_2 + \widecheck{\ka}_2 - \frac{t_1 t_2}{2} \right) \right] , \\
    \widebar{K}_{2,-2} = & \, \ka_0 \left[ \widebar{t}_{2,-1} - \frac{t_1^2}{4} + \frac{\ka_0}{2}\left( \widecheck{k}_2 + \widecheck{\ka}_2 \right) \right] ,
\end{align}
and the expression for $\widebar{T}_{2,-2}$ is extremely long and uninformative and we do not display it. As expected, the integrating constants $\widebar{t}_i$ can be freely chosen. With this last calculation we completely specified the transformation theory at second order in the spin.
\end{widetext}

\bibliography{literature}

\end{document}